\pdfoutput=1

\documentclass[times,authoryear,1p,nopreprintline,12pt]{elsarticle}

\usepackage{framed,multirow,fullpage}

\usepackage{amssymb}
\usepackage{mathtools}
\usepackage{latexsym}
\usepackage{microtype}

\usepackage{url}
\usepackage{xcolor}
\definecolor{newcolor}{rgb}{.8,.349,.1}

\usepackage{bm,tikz,upgreek}

\newcommand\figref[1]{Figure \ref{#1}}

\newcommand\chref[1]{\textcolor{red}{REPLACE THIS}}

\newcommand\from{\leftarrow}

\newcommand\p{\partial}
\newcommand\gr{\bm{\nabla}}

\renewcommand\vec[1]{\mathbf{#1}}

\newcommand\vc{\vec{c}}
\newcommand\vx{\vec{x}}
\newcommand\vX{\vec{X}}

\newcommand\vv{\vec{v}}

\newcommand\vw{\vec{w}}
\newcommand\vq{\vec{q}}
\newcommand\vpsi{\bm{\uppsi}}

\newcommand\ten[1]{\vec{#1}}
\newcommand\ts{\bm{\upsigma}}
\newcommand\ttau{\bm{\uptau}}
\newcommand\tI{\ten{I}}
\newcommand\tD{\ten{D}}

\newcommand\nhat{\hat{\vec{n}}}
\newcommand\xhat{\hat{\vec{x}}}
\newcommand\yhat{\hat{\vec{y}}}

\newcommand\f{\text{f}}
\newcommand\s{\text{s}}

\newcommand\vvs{\vec{v}_\s}
\newcommand\vvf{\vec{v}_\f}
\newcommand{\trans}{\mathsf{T}}

\newcommand\vf{\vec{f}}

\usepackage{hyperref}
\hypersetup{colorlinks,breaklinks,
            urlcolor=Teal,
            linkcolor=MediumBlue,
            citecolor=NavyBlue,
            pdfauthor={Nicholas J. Derr and Chris H. Rycroft},
            pdftitle={A projection method for porous media flow}
}

\DeclareMathOperator{\Det}{det}
\DeclareMathOperator{\Dev}{dev}
\DeclareMathOperator{\Tr}{tr}

\begin{document}

\begin{frontmatter}

\title{A projection method for porous media flow}%

\author[1]{Nicholas J. Derr\corref{cor1}}
\author[1,2]{Chris H. Rycroft\corref{cor2}}
\cortext[cor1]{Email: nicholas.derr@gmail.com}
\cortext[cor2]{Email: chris.rycroft@gmail.com}

\address[1]{John A. Paulson School of Engineering and Applied Sciences, Harvard University, 29 Oxford Street, Cambridge, MA 02138}
\address[2]{Mathematics Group, Lawrence Berkeley National Laboratory, 1 Cyclotron Road, Berkeley, CA 94720}

\begin{abstract}
Flow through porous, elastically deforming media is present in a variety of natural contexts ranging from large-scale geophysics to cellular biology. In the case of incompressible constituents, the porefluid pressure acts as a Lagrange multiplier to satisfy the resulting constraint on fluid divergence. The resulting system of equations is a possibly non-linear saddle-point problem and difficult to solve numerically, requiring nonlinear implicit solvers or flux-splitting methods. Here, we present a method for the simulation of flow through porous media and its coupled elastic deformation. The pore pressure field is calculated at each time step by correcting trial velocities in a manner similar to Chorin projection methods. We demonstrate the method's second-order convergence in space and time and show its application to phase separating neo-Hookean gels.
\end{abstract}


\end{frontmatter}

\section{Introduction}

Deforming porous media appear across nature, in settings such as soil consolidation \citep{biot1941}, river bed and channel formation \citep{rodriguez-iturbe1997,abrams_growth_2009}, the planet's mantle \citep{mckenzie_generation_1984,spiegelman_flow_1993}, glacial melt runoff \citep{hewitt2011}, living tissue \citep{ehlers_advances_2010,ranft_tissue_2012}, cellular interiors \citep{dembo_mechanics_1989,howard2002,bendix_quantitative_2008,schaller_crosslinking_2013}, and simple organisms \citep{radszuweit_active_2014}.
The essential similarity between all of these cases are the governing behaviors laid out in the poroelasticity theory introduced by \citet{biot1941}, wherein a porous solid skeleton is suffused with fluid.
Contraction of the skeleton at a point in space raises the pore pressure, which gradually equilibriates via fluid exchange with neighboring pores, leading to diffusion of the pressure perturbation throughout the medium.
\citet{rice1976} categorized the limiting behaviors of this response as \textit{undrained} (on time scales faster than fluid exchange) and \textit{drained} (on times long enough for fluid exchange to have taken place.)
In contexts such as polymer gels, the fluid and solid constituents are both effectively incompressible \citep{doi_gel_2009}, such that fluid exchange can be taken as effectively instantaneous and the system behavior is effectively drained at all times.

Even in the linear regime,
solving the coupled set of equations for pressure and displacements is non-trivial.
One widely used method for doing so is the \textit{fixed-stress splitting method} \citep{kim2011,dana2022}, where the flow and mechanical problems are solved in succession.
This is convenient in that it allows separate code bases to be applied to individual parts of the problem.
However, this arrangement typically corresponds to obtaining solutions on separate grids, necessitating coupling procedures and interpolation between grids.

Many of the biological examples of porous media---especially polymer gels---are capable of undergoing large volume changes that correspond to large deformation and require non-linear constitutive models \citep{bendix_quantitative_2008,doi_gel_2009}.
In this regime, the equations of motion can be cast in either of two different perspectives: Lagrangian, where variables are tracked at a given material point; or Eulerian, where they are tracked at a given point in space.
Large deformation theories typically use a Lagrangian  formulation for the porestructure, tracking the displacement and fluid flux (relative to solid motion) at each material point \citep{deboer2005}.
Alternatively, equivalent Eulerian relations for the system can be derived from mixture theory as shown by \citet{coussy_mixture_1998}, tracking the density and momentum flux of each phase at each point in space.
These methods have previously been used to investigate porous media flow as well \citep{spiegelman_simple_1987}, relying on simple geometries for analytical solutions.

In this paper, we propose a fully Eulerian method for porous media flow using this latter formulation.
While the method is general, we first intend to apply it to the study of biological polymer gels.
Individual monomers are significantly less compressible than grains from rock or soil mechanics \citep{howard2002}, leading to the assumption above of constituent incompressibility.
This constraint is satisfied at all times by the pore pressure field, acting as a Lagrange multiplier in much the same way as pressure in the Navier--Stokes equations \citep{deboer2001}.
This corresponds to a non-linear multi-component saddle-point system which would need to be solved several times per timestep, requiring specialized preconditioners for the pressure block analogous to those required for Brinkman flow \citep{borrvall2003}.

Instead, we satisfy the gel's incompressibility constraint by adapting the projection method for Navier--Stokes simulation, originally developed by \citet{chorin67,chorin68}. 
Here, we effectively apply a projection method \citep{bell89,almgren96,puckett97,brown2001} to the time-integration of the total mean velocity field obtained by volume averaging those of each constituent.
However, we do not use this mean velocity field as a primary simulation variable as in the Lagrangian formulation above.
At each time step, an intermediate mean velocity field is calculated in the absence of the constraint.
This function is projected onto the space of incompressible vector fields by solving a Poisson relation, and the result is used to correct each constituent velocity field so that the incompressibility constraint is satisfied.

In the next section, we review a derivation of the equations of porous media flow, following the approach of \citet{coussy_mixture_1998}. In section 3, we describe the second-order spatial dicretizations used to approximate derivatives and update rules for time-integration. In section 4, we apply the method to two test cases: a manufactured solution for confirming the spatial and temporal rates of convergence, and phase separation of viscoelastic polymer gels.

\section{Continuum theory}
Consider a solid structure of characteristic pore size $\xi$ filled with permeating fluid. We define continuum fields following the approach of \citet{coussy_mixture_1998}.
On lengths comparable to the pore size---the \textit{microscale}---each point in space is occupied by polymer or solvent and has associated with it a single value for fields such as density, momentum, or Cauchy stress state.
On larger lengths $L \gg \xi$---the \textit{macroscale}---we can define corresponding fields at a position $\vx$ on the macroscale by averaging over a microscale volume $V_l = \{\vx_\mu : |\vx_\mu - \vx| < l\}$ centered at $\vx$ and of characteristic size $l$ with $L \gg l \gtrsim \xi$. In particular, we can perform this averaging procedure for a single phase, identified by an index $\alpha$, as
\begin{equation}
	\left<\psi\right>_\alpha \left(\vx,t\right) = \int_{V_l} \psi_\mu(\vx_\mu) K(\vx_\mu - \vx) I_\alpha(\vx_\mu) \ dV.
\end{equation}
Here $\psi_\mu$ is a tensor or scalar field defined at the microscale, $K$ is a normalized kernel defined over balls of radius $l$, and $I_\alpha$ is an indicator function equal to 1 if the point of space at $x_i^\mu$ is occupied by constituent $\alpha$ and 0 otherwise.
We consider for each phase $\alpha$ a volume fraction $\phi_\alpha$, density $\rho_\alpha$, velocity $\vv_\alpha$, and Cauchy stress tensor $\ttau_\alpha$ defined according to
	\begin{equation}
		\phi_\alpha = \left<1\right>_\alpha, \qquad
		\rho_\alpha = \left<\rho_\mu\right>_\alpha, \qquad
		\rho_\alpha \vv_\alpha = \left<\rho_\mu \vv_\mu\right>_\alpha, \qquad
		\ttau_\alpha = \left<\ttau_\mu\right>_\alpha.
	\end{equation}
Importantly, the densities defined here are \textit{apparent}---corresponding to mass per unit volume of total material rather than volume of constituent $\alpha$---and the corresponding \textit{intrinsic} densities can be written \mbox{$\hat\rho_\alpha = \rho_\alpha/\phi_\alpha$}.

In the same way, we have defined apparent and intrinsic stress tensors.
In small-deformation regimes where $\phi$ is approximately constant, it is common to close the problem by specifying the intrinsic stress $\hat{\ttau}_{\alpha} = \ttau_\alpha / \phi_\alpha$ as a function of the system variables \citep{radszuweit_active_2014,weber_differential_2018}.
However, large deformation theories specify a form of the apparent stress tensor based on the energetic cost of large-scale solid phase deformation \citep{deboer2005}.
Derivatives and integrals can be defined in terms of similar averaging procedures \citep{coussy_mixture_1998}.
From this point forward, we will concern ourselves only with relations at the macroscale.

In this work we will consider biphasic mixtures such that one phase is solid- and the other fluid-like, in the sense of stresses depending on deformation and deformation rate, respectively.
Accordingly, we let $\smash{\alpha=\text{f}}$ or s.
Since $\phi_\f + \phi_\s = 1$ by definition, the volume fractions of both phases represent a single degree of freedom; as such, we will frequently use $\phi = \phi_\s$ to simply denote the solid volume fraction.

\subsection{Governing equations}
At this point, we apply the principles of conservation of mass and conservation of momentum to each phase in order to calculate the equations of motion associated with gels \citep{doi_gel_2009}.
Conservation of mass within each phase gives
\begin{equation}
	\frac{\p \rho_\alpha}{\p t} + \gr \cdot \left(\rho_\alpha \vv_\alpha\right)= s_{\alpha\from\beta} + s_\alpha,
\end{equation}
where $s_{\alpha\from\beta}$ for $\alpha \ne \beta$ is the rate at which mass of phase $\beta$ is transformed into mass of phase $\alpha$ and $s_\alpha$ is the source rate of mass of phase $\alpha$.
Conservation of mass requires $s_{\alpha\from\beta}+s_{\beta\from\alpha}=0$.
If the constituents are incompressible such that the intrinsic density $\hat\rho_\alpha$ is constant for each phase, this simplifies to a set of volume fraction conservation equations,
\begin{equation}
	\frac{\p \phi_\alpha}{\p t} + \gr \cdot \left(\phi_\alpha \vv_\alpha\right)= \frac{s_{\alpha\from\beta}+s_\alpha}{\hat{\rho}_\alpha}.
\label{eq:mass_conv_phase}
\end{equation}
With substitution, \eqref{eq:mass_conv_phase} can be equivalently represented by a single volume fraction conservation equation and an incompressibility condition on the mean material velocity,
\begin{subequations}
	\begin{equation}
		\frac{\p \phi}{\p t} + \gr \cdot \left(\phi \vvs\right)=\frac{s_{\s\from\f}}{\hat{\rho}^\s}, \label{eq:mass_con}
	\end{equation}
\begin{equation}
	\gr \cdot \left[\phi \vvs + (1-\phi)\vvf\right] = \left(\frac{1}{\hat{\rho}^\s} - \frac{1}{\hat{\rho}^\f}\right) s_{\s\from\f} + \frac{s_\s}{\hat{\rho}^\s} + \frac{s_\f}{\hat{\rho}^\f}.
\label{eq:incompr}
\end{equation}\label{eq:mass_incompr}\end{subequations}%
The first term on the right-hand side represents volume change due to phase change (e.g.\@ freezing or melting).
The constraint \eqref{eq:incompr} necessitates the introduction of a macroscopic variable $p$ to enforce it in the manner of a Lagrange multiplier, exactly as in the case of the incompressible Navier--Stokes equations.
Going forward, we assume no material sources or phase change, so that $s_{\s\leftarrow \f} = s_\s = s_\f = 0$ and
\begin{equation}
\gr \cdot \left[\phi \vvs + (1-\phi) \vvf\right] = 0.
\end{equation}

Performing the same analysis using conservation of momentum rather than mass, we
obtain a momentum balance equation
\begin{equation}
	\rho_\alpha \left(\frac{\p \vv_\alpha}{\p t} + \vv_\alpha \cdot \gr \vv_\alpha\right) = \gr \cdot\ttau_\alpha + \vf_{\alpha\from\beta} + \vf_\alpha, \label{eq:mom_balance}
\end{equation}
	where $\vf_\alpha$ is a body force acting only on phase $\alpha$ and $\vf_{\alpha\from\beta} + \vf_{\beta\from\alpha} = \vec{0}$ are a pair of equal-and-opposite forces exerted one each phase by the other.
	These originate from integrals of surface tractions over phase boundaries on the microscale. Assuming the existence of an isotropic stress $-p\tI$ everywhere on the microscale, it can be shown \citep{coussy_mixture_1998} that there exists a pressure contribution to $\vf_{\alpha\from\beta}$ equal to $p \gr \phi_\alpha$.
We assume the remainder of the interaction force stems from the viscous stresses applied to solid surface by flow of the penetrating fluid through the porestructure.
Accordingly, we let
\begin{equation}
	\vf_{\alpha\from\beta} = p\partial_i \phi_\alpha + \Gamma(\phi) \left(\vv_\beta - \vv_\alpha\right),
\end{equation}
where $\Gamma(\phi) \sim \mu /\xi^2$ is a prefactor scaling with pore-scale viscous forces.
We will let $\Gamma = \Gamma_0 \phi^2/(1-\phi)$, $\Gamma_0$ a constant, for consistency with the Carman--Kozeny relation describing flow through packed spheres \citep{carman1937,kozeny1927}.

To reflect the effect of letting $p$ exist in both phases, we decompose the Cauchy stress $\ttau_\alpha$. We define the effective stress $\ts_\alpha = \ttau_\alpha + p \tI$ as the portion of macroscale stress not originating from pressure.
We assume there are no other sources of stress within the fluid phase besides the  pressure $p$ and so set $\ts_\f = \vec{0}$, i.e.\@ $\ttau_\f = -p \tI$.
This yields the identification of the pressure $p$ as the porefluid pressure as in the theory of \citet{biot1941}.
Finally, whatever the eventual definition of $\ts_\s$, we assume the characteristic stress magnitudes are large relative to the inertial forces on the left-hand side of \eqref{eq:mom_balance}, such that the left-hand side can be omitted. In practice, we will preserve the partial time derivative and neglect the advective term, reasoning that it is dominated by the right-hand side terms.

With this notation and these assumptions, the set of model equations are
\begin{subequations}
	\begin{equation}
		\frac{\p\phi}{\p t} + \gr\cdot (\phi \vvs) = 0, \label{eq:gel:phi}
	\end{equation}
	\begin{equation}
		\rho_\s \frac{\p \vvs}{\p t} \approx \vec{0} = \gr \cdot\ts_\s -\phi  \gr p + \Gamma(\phi)\left(\vvf-\vvs\right), \label{eq:gel:vs}
	\end{equation}
	\begin{equation}
		\rho_\f \frac{\p \vvf}{\p t} \approx \vec{0} = -(1-\phi) \gr p - \Gamma(\phi) \left(\vvf - \vvs\right), \label{eq:gel:vf}
	\end{equation}
	\begin{equation}
		\gr \cdot \left[\phi \vvs + (1-\phi) \vvf\right] = 0. \label{eq:gel:p}
	\end{equation}\label{eq:gel}\end{subequations}%
Note the two momentum balance equations can be summed to obtain a total material mechanical equilibrium,
	\begin{equation}
		\vec{0} = \gr \cdot \left(-p\tI +\ts_\s\right) + \vf_\s + \vf_\f,
	\end{equation}
	and in the absence of a body force $\vf_\f$,  the fluid momentum equation can be written in terms of the volumetric flux $\vq = (1-\phi) \left(\vvf-\vvs\right)$ as a Darcy-type relation
	\begin{equation}
		\vq = -\frac{(1-\phi)^3}{\Gamma_0 \phi^2} \gr p,
	\end{equation}
	where we have substituted for $\Gamma(\phi)$, obtaining the usual Carmen--Kozeny proportionality between flow rate and pressure gradients.

\section{Numerical method}
We now describe the numerical method by which we solve \eqref{eq:gel}.
We begin by discussing the discretization of our domain into a periodic, two-dimensional grid.
While we use the language of two dimensions to report the form of finite difference stencils, we also consider their form in three dimensions to allow for use of the method in either context.

\subsection{Grid discretization}
In two dimensions, we discretize a rectangular domain $[a_x,b_x] \times [a_y, b_y] \in \mathbb{R}^2$ into an $m\times n$ grid of cells $\Omega_{i,j}$ of width $\Delta x$ and height $\Delta y$.
The left bottom corner of cell $\Omega_{i,j}$ is
\begin{equation}
	\vc_{i,j} = \left(a_x + i\Delta x\right)\xhat + \left(a_y + j\Delta y\right)\yhat,
\end{equation}
its center is
\begin{equation}
	\vx_{i,j} = \vc_{i,j} + \left(\frac{\Delta x \ \xhat + \Delta y \ \yhat}{2}\right),
\end{equation}
and the center of its left and bottom walls are given by
\begin{equation}
	\vw_{i,j}^l = \vc_{i,j} + \frac{\Delta y \ \yhat}{2} ,  \qquad \vw_{i,j}^d = \vc_{i,j} + \frac{\Delta x \ \xhat}{2}.
\end{equation}
We consider of series of time intervals $\Delta t$, such that we may write the time of one step of the simulation as $t_n = n\Delta t$.

We discretize the simulation fields such that each cell is associated with a single value of the solid fraction and phase velocities located at its center,
\begin{equation}
	\phi_{i,j}^n = \phi(\vx_{i,j},t_n), \qquad (\vvs)_{i,j}^n = \vvs(\vx_{i,j},t_n), \qquad (\vvf)_{i,j}^n = \vvf(\vx_{i,j},t_n),
\end{equation}
Pressure and stress fields are located on cell corners and walls, respectively, such that in each cell we have
\begin{equation}
	p_{i,j}^n = p(\vc_{i,j},t_n), \qquad \ts_{i,j}^{l,n} = \ts(\vw_{i,j}^l,t_n), \qquad \ts_{i,j}^{d,n} = \ts(\vw_{i,j}^d,t_n).
\end{equation}

This allows for the simple calculation of finite difference approximations to field derivatives.
In particular, the pressure gradient at cell centers is calculated
\begin{equation}
	\left(\gr p\right)^n_{i,j} = \left[\frac{p^n_{i+1,j+1}+p^n_{i+1,j}-p^n_{i,j+1}-p^n_{i,j}}{2\Delta x}\right]\xhat  + \left[\frac{p^n_{i+1,j+1}+p^n_{i,j+1}-p^n_{i+1,j}-p^n_{i,j}}{2\Delta y}\right]\yhat,
\end{equation}
and the stress divergence at cell centers is
\begin{equation}
	\left(\gr \cdot \ts\right)^n_{i,j} = \frac{\xhat \cdot \ts^{l,n}_{i+1,j} - \xhat \cdot \ts^{l,n}_{i,j}}{\Delta x} + \frac{\yhat \cdot \ts^{d,n}_{i,j+1} - \yhat \cdot \ts^{d,n}_{i,j}}{\Delta y}.
\end{equation}
We also calculate the solid flux divergence at cell centers first by introducing its value on the left and bottom walls,
\begin{subequations}
\begin{equation}
	\left(\phi v_s\right)^{l,n}_{i,j} = \frac{\left(\phi^n_{i-1,j}+\phi^n_{i,j}\right)\left(\xhat \cdot (\vvs)^n_{i-1,j} + \xhat \cdot (\vvs)^n_{i,j}\right)}4, \nonumber \\
\end{equation}
\begin{equation}
	\left(\phi v_s\right)^{d,n}_{i,j} = \frac{\left(\phi^n_{i,j-1}+\phi^n_{i,j}\right)\left(\yhat \cdot (\vvs)^n_{i,j-1} + \yhat \cdot (\vvs)^n_{i,j}\right)}4,
\end{equation}
\end{subequations}
and then calculating the divergence analogously to the stress,
\begin{equation}
  \left(\gr \cdot (\phi \vvs\right))^n_{i,j} = \frac{\left(\phi v_s\right)^{l,n}_{i+1,j} - \left(\phi v_s\right)^{l,n}_{i,j}}{\Delta x} + \frac{\left(\phi v_s\right)^{d,n}_{i,j+1} - \left(\phi v_s\right)^{d,n}_{i,j}}{\Delta y}.
\end{equation}

\subsection{Arbitrary finite difference expressions}
Now, we introduce second-order spatially converging finite difference expressions for calculating discretized representations of various differential operators.
We will explicitly write out the finite difference rules in two dimensions, letting $\psi_{i,j} = \psi(\vx_{i,j})$ denote an arbitrary cell-centered scalar field and $\vpsi_{i,j} = \vpsi(\vx_{i,j})$ an arbitrary vector field, with $\vpsi = \psi_x \xhat + \psi_y \yhat$. 

First, we discuss the calculation of gradients of cell-centered fields on cell walls.
Writing the two-dimensional gradient out in terms of its components, $\left(\gr \psi\right)^w = (\partial_x \psi)^w \xhat +(\partial_y \psi)^w\yhat$, we will introduce separate finite difference rules for the normal derivatives $(\partial_x\psi)^l$ and $(\partial_y\psi)^d$ and the tangential derivatives $(\partial_y\psi)^l$ and $(\partial_x\psi)^d$.
 The normal derivatives across the wall can be calculated by a simple difference as
\begin{equation}
    (\partial_x \psi)^l = \frac{\psi_{i,j}-\psi_{i-1,j}}{\Delta x}, \qquad (\partial_y \psi)^d = \frac{\psi_{i,j} - \psi_{i,j-1}}{\Delta y}.
\end{equation}
We compute tangential derivatives by averaging the differentiated quantity on the cell walls above and below the gradient location in the derivative's direction, i.e.
\begin{subequations}
\begin{equation}
	\left(\partial_y \psi\right)^l = \frac{\psi_{i,j+1}+\psi_{i-1,j+1} - \psi_{i,j-1} - \psi_{i-1,j-1}}{4\Delta y},
\end{equation}
	\begin{equation}
	\left(\partial_x \psi\right)^d =\frac{\psi_{i+1,j} + \psi_{i+1,j-1} - \psi_{i-1,j} - \psi_{i-1,j-1}}{4\Delta x}.
\end{equation}
\end{subequations}
While we have written these expressions in terms of an arbitrary scalar field, they also give the finite difference rules for the calculation of gradients of vectors.
In particular, for $(\gr \vpsi)^w = \xhat (\partial_x \vpsi)^w + \yhat (\partial_y \vpsi)^w$, the rules for the calculation of $(\partial_x \bm \psi)^w$ and $(\partial_y \bm \psi)^w$ are exactly as above, with the vector field $\vpsi$ substituted for the scalar field $\psi$.
This implies we identify the row of a gradient tensor with the derivative index unless otherwise specified.

The finite difference expressions can be generalized to three dimensions. In two dimensions each wall features one normal and one tangential derivative, but in three dimensions each wall features one normal and two tangential derivatives.

Now we discuss the application of the advective derivative $\vv_\alpha \cdot \gr$ using an essentially non-oscillatory (ENO) method which is stable in the face of shocks in the differentiated fields.
We use the second-order ENO2 method \citep{osher88,osher1991}.
In brief, the derivative in each direction is calculated from a three-node stencil drawn from a set of five possible values.
The three selected nodes are chosen so that the approximated derivative is either centered or upwinded and so that the nodes feature the smoothest variation of the differentiated field in terms of the second derivative magnitude.

For simplicity, we consider a one-dimensional scalar field $\psi$ and scalar velocity $v$ located at nodes spaced a distance $h$ apart in the $x$-direction. The ENO2 approximation to the quantity $v d\psi/dx$ is calculated as follows.
At one node, we consider the set of five values $S = \{\psi_{-2},\psi_{-1},\psi_0,\psi_1,\psi_2\}$ corresponding to the field's value at that cell and its first two neighbors in direction.
With these values, we can calculate the quantities
\begin{equation}
	s_\text{b} = |\psi_{-2} - 2\psi_{-1} + \psi_0|, \qquad s_\text{c} = |\psi_{-1} - 2\psi_0 + \psi_1|, \qquad s_\text{f} = |\psi_{0} - 2\psi_1 + \psi_2|,
\end{equation}
which are proportional to the absolute magnitude of the second derivative at the center cell as calculated using backward, centered, and forward second-order stencils.
We will use these as a proxy for which set of three nodes feature sharper variation of the function $\psi$.
We can also calculate three approximations to the first derivative $f$ using each set of three nodes,
\begin{equation}
	f_\text{b} = \frac{\psi_{-2} -4\psi_{-1} + 3\psi_0}{ 2h}, \qquad f_\text{c} = \frac{\psi_1 - \psi_{-1}}{2h}, \qquad f_\text{f} = \frac{-3\psi_0 + 4 \psi_1 - \psi_2}{2h}.
\end{equation}
The approximation chosen for the first derivative is as follows: it uses either the centered or upwinded stencil, choosing between the two based on which has a lower sharpness as defined above.
In other words, denoting the ENO2 approximation to the first derivative as $E[S;v,h]$, we can write
\begin{equation}
	E\left[S;v,h\right]= \begin{cases} f_\text{b}, & v > 0 \text{ and } s_\text{b} < s_\text{c},\\ f_\text{f}, & v < 0 \text{ and } s_\text{f} < s_\text{c}, \\ f_\text{c}, & \text{otherwise.} \end{cases}
\end{equation}
Now, we explicitly relate this one-dimensional rule to the $D$-dimensional vector operator applied to scalars and vectors in two dimensions.
Consider a scalar field discretized at cell centers such that $\psi_{i,j} = \psi(\vx_{i,j})$.
If we denote $v_x = \xhat \cdot \vv_{i,j}$, $v_y = \yhat \cdot \vv_{i,j}$, then we can introduce the set of five nodes in each dimension necessary for applying the ENO method,
\begin{equation}
	S^{(x)}_{i,j} = \{\psi_{i-2,j},\psi_{i-1,j},\psi_{i,j},\psi_{i+1,j},\psi_{i+2,j}\}, \qquad S^{(y)}_{i,j} = \{\psi_{i,j-2},\psi_{i,j-1},\psi_{i,j},\psi_{i,j+1},\psi_{i,j+2}\},
\end{equation}
and write the scalar advective derivative
\begin{equation}
	\left(\vv \cdot \gr \psi\right)_{i,j} = v_x E\left[S^{(x)}_{i,j};v_x,\Delta x\right]  + v_y E\left[S^{(y)}_{i,j};v_y,\Delta y\right].
\end{equation}
The application to vector fields corresponds to independently applying the operator to each scalar component field, i.e.
for a vector field $\vpsi = \psi_x \xhat + \psi_y \yhat$, we write
\begin{equation}
	\left(\vv \cdot \gr \vpsi\right)_{i,j} = \left(\vv \cdot \gr \psi_x\right)_{i,j} \xhat + \left(\vv \cdot \gr \psi_y\right)_{i,j} \yhat.
\end{equation}
This differs from the gradient difference rule in that different stencils may be chosen for the derivative of each component field, so it cannot be written as a simple difference of vector values.

\subsection{Update rules}
Relations \eqref{eq:gel:phi}--\eqref{eq:gel:vf} each contain a partial time derivative, and as such are amenable to explicit timestepping, but the incompressibility condition \eqref{eq:gel:p} contains no such update rule for the pressure.
As previously mentioned, we adapt Chorin-style projection methods to the system of equations presented here \citep{chorin67,chorin68}.
We calculate intermediate velocity fields $\vvs^*$ and $\vvf^*$ using the update rules below that do not in general satisfy the incompressibility condition \eqref{eq:gel:p}.
Then, an elliptic PDE is solved to calculate the pressure required to enforce incompressibility during the period $t \in [t_n,t_{n+1}]$. As such, the pressure obtained can be identified as that as the interval's midpoint, i.e.\@ $p^{\smash{n+\frac12}}_{i,j} = p(\vc_{ij},t_n + \frac12 \Delta t)$, as detailed in e.g.\@ by \citet{brown2001}.
We include the previous time-step's pressure $p^{\smash{n-\frac12}}_{i,j}$ in the calculation of the intermediate velocities, so that the elliptic PDE correpsonds to the pressure increment $q^n_{ij} = p^{\smash{n+\frac12}}_{ij} - p^{\smash{n-\frac12}}_{ij}$.

We will decompose the stress tensor into elastic and viscous parts $\ts_\s = \ts_e + \ts_v$.
The portion of the velocity update corresponding to the viscous stress will be completed with the semi-implicit Crank--Nicolson method \cite{crank47}.
Every other aspect of the problem is explicitly integreated using the second-order Heun's method, a two-stage explicit Runge--Kutta method. We let $\overline{n+1}$ denote an approximate value at the next-time step computed with a forward Euler update. An arbitrary field $\psi$ with $\p\psi/\p t = g(\psi,t)$ would thus satisfy
\begin{equation}
	\psi^{\overline{n+1}} = \psi^n + \Delta t g(\psi,t_n),
\end{equation}
and a complete update with Heun's method would give
\begin{equation}
	\psi^{n+1} = \psi^n + \frac{\Delta t}{2}\left[g(\psi^n,t_n) + g(\psi^{\overline{n+1}},t_{n+1})\right].
\end{equation}
Thus, the solid fraction $\phi$ satisfies
\begin{equation}
	\frac{\phi^{n+1}_{i,j} - \phi^n_{i,j}}{\Delta t} = -\frac12 \left[\left(\gr\cdot \phi \vvs\right)^n_{i,j} + \left(\gr\cdot \phi \vvs\right)^{\overline{n+1}}_{i,j}\right].
\end{equation}
Letting the drag force in cell $\Omega_{i,j}$ be denoted $\bm{d}^n = \Gamma\left(\phi_{i,j}^n)((\vvf)_{i,j}^n - (\vvs)_{i,j}^n\right)$, we introduce solid and fluid velocity updates
\begin{multline}
  \frac{(\vvs)^*_{i,j} -(\vvs)^n_{i,j}}{\Delta t} = \frac12\left[\frac{\left(\gr \cdot \ts_e\right)^n_{i,j} + \bm d^n}{\hat{\rho}_\s\phi_{i,j}^n} + \frac{\left(\gr \cdot \ts_e\right)^{\overline{n+1}}_{i,j} + \bm d^{\overline{n+1}}}{\hat{\rho}_\s\phi_{i,j}^{\overline{n+1}}}\right] 
	 \\ - \frac{\left(\gr p\right)^{n-\frac12}_{i,j}}{\hat{\rho}_s} + \frac12\left[\frac{\left(\gr \cdot \ts_v\right)^n_{i,j}}{\hat{\rho}_\s \phi^n_{i,j}} + \frac{\left(\gr \cdot \ts_v\right)^{n+1}_{i,j}}{\hat{\rho}_\s \phi^{n+1}_{i,j}}\right], \label{eq:gel:inter_s}
\end{multline}
\begin{equation}
	\frac{(\vvf)^*_{i,j}-(\vvf)^n_{i,j}}{\Delta t} = \frac12\left[\frac{-\bm{d}^n}{\hat{\rho}_f (1-\phi^n_{i,j})} + \frac{-\bm{d}^{\overline{n+1}}}{\hat{\rho}_f (1-\phi^{\overline{n+1}}_{i,j})}\right] - \frac{\left(\gr p\right)^{n-\frac12}_{i,j}}{\hat{\rho}_f},
\end{equation}
where $\vvs^*$ and $\vvf^*$ are a set of intermediate velocities that we do not expect to satisfy the condition of incompressibility.
From these we can compute an intermediate material velocity
\begin{equation}
	\bar{\vec{V}}^*_{i,j} = \phi^*_{i,j} (\vvs)^*_{i,j} + (1-\phi^*_{i,j})(\vvf)^*_{i,j}.
\end{equation}
At this point, we solve for the correction field $q^n_{i,j}$ such that $p_{i,j}^{\smash{n+\frac12}} = p^{\smash{n-\frac12}}_{i,j} + q^{n}_{i,j}$.
For ease of notation, we introduce the specific volumes $\nu_\alpha = 1/\hat{\rho}_\alpha$. To indicate the following analysis is without reference to any particular discretization, we omit the $(i,j)$ subscripts.
We write the final velocity correction as
\begin{equation}
	\frac{\vvs^{n+1} -\vvs^*}{\Delta t} = -\nu_\s\gr q^{n}, \qquad \frac{\vvf^{n+1}-\vvf^*}{\Delta t} = -\nu_\f\gr q^{n},
\end{equation}
The weighted average velocity $\bar{\vec{V}} = \phi \bm u + (1-\phi) \bm v$ at timestep $n+1$ must be divergence-free.
We construct the term $\gr \cdot \bar{\vec{V}}^{n+1}$ by multiplying each of the above relations by its corresponding phase fraction, applying the divergence operator, and adding the two.
Setting it to zero in accordance with the incompressibility condition yields an elliptic PDE for the pressure increment $q$,
\begin{equation}
	\gr \cdot \left[\Delta t \bar\nu(\phi) \gr q^n \right] = \gr \cdot \bar{\vec{V}}^*, \label{eq:q}
\end{equation}
where we have introduced  \mbox{$\bar\nu(\phi) = \phi \nu_\s + (1-\phi)\nu_\f$}, the material specific volume.

Note the approximate $(n+1)$ velocity field in the Heun's method time updates correspond to the intermediate velocities, not the corrected divergence-free velocities, introducing a local error $e^n \propto \Delta t (\vv^{n+1}_\alpha-\vv^*_\alpha)$ at each timestep. However, this difference scales as $(\vv^{n+1}_\alpha -\vv^*_\alpha) \propto \Delta t \gr q^n$, and the pressure increment $q^n$ scales with $\Delta t$ as well \citep{brown2001}. Thus, $e^n \propto \Delta t^3$ and the second-order convergence in time is preserved.

\subsection{Numerical solution of the elliptic problem}
Finally, we discuss the solution of \eqref{eq:q} using the finite element method, representing the pressure correction as a superposition of basis functions $q(\vx,t_n) = q_{i,j}^{n} \varphi_{i,j}(\vx)$. 
On each cell $\Omega_{i,j}$, we introduce four basis functions.
Letting $c_x = \xhat\cdot\vc_{i,j}$ and $c_y = \yhat \cdot \vc_{i,j}$, and letting $\xi = (x-c_x)/\Delta x$, $\eta = (y - c_y)/\Delta y$,
\begin{equation}
	\varphi^{\text{dl}}_{i,j} = (1-\xi)(1-\eta), \ \varphi^{\text{dr}_{i,j}} = \xi (1-\eta), \ \varphi^\text{ul}_{i,j} = (1-\xi)\eta, \ \varphi^\text{ur}_{i,j} = \xi \eta,
\end{equation}
such that each basis function takes the value 1 at one corner of the cell and 0 at each of the others.
We can then define the global basis function as
\begin{equation}
	\varphi_{i,j}(\vx) = \begin{cases}
		\varphi^\text{dl}, & \vx \in \Omega_{i,j},\\
		\varphi^\text{dr}, & \vx \in \Omega_{i-1,j},\\
		\varphi^\text{ul}, & \vx \in \Omega_{i,j-1},\\
		\varphi^\text{ur}, & \vx \in \Omega_{i-1,j-1},\\
		0, & \text{otherwise.}
	\end{cases}
\end{equation}
Multiplying by a test function $Q$ and integrating by parts, we obtain the weak form of \eqref{eq:q},
\begin{equation}
	\int_V \Delta t \bar\nu(\phi) \gr Q \cdot \gr q \ dV = \int_V \bar{\vec{V}}^*\cdot \gr Q \ dV - \int_{\partial V} Q \nhat \cdot\left(\bar{\vec{V}}^* - \Delta t \bar\nu(\phi) \gr q\right)dS,
\end{equation}
but the surface integral on the right-hand side vanishes due to the periodicity of our domain.

Letting a Greek index denote a single corner node of the grid, we write the test function $Q = Q_\nu \varphi_\nu(\vx)$ and correction field $q=q_\nu \varphi_\nu(\vx)$ in terms of our basis functions.
The above then corresponds to a linear system of equations determining the $q$-coefficients,
\begin{equation}
	M_{\nu\gamma} q_\gamma  = b_\nu, \label{eq:proj}
\end{equation}
with the operator $M$ and source vector $b$ defined as
\begin{equation}
	M_{\nu\gamma} = \int_V \Delta t \bar \nu(\phi) \gr \varphi_\alpha \cdot \gr \varphi_\beta dV, \qquad b_\alpha = \int_V \bar{\vec{V}}\cdot \gr \varphi_\alpha dV.
\end{equation}

At this point, we consider a model system in order to test the method. After defining a stress tensor in the following section, we will be able to specify a form for $\left(\gr \cdot \ts\right)^n_{i,j}$ in \eqref{eq:gel:inter_s}, which is the only undefined piece of information left to specify in the steps above.

\paragraph{Stress specification}
As mentioned, we decompose the stress tensor into elastic and viscous parts $\ts = \ts_e + \ts_v$.
Elastic stresses are produced by deformation of a solid body occupying the domain $\Omega$ relative to an unstressed reference configuration $\Omega_0$.
We will calculate elastic stresses using the reference map technique (RMT) \citep{kamrin12,rycroft20,lin2022}.
A position in the reference frame $\bm \xi \in \Omega_0$ corresponds to a particular material point of the deforming body, while a position in the deformed frame $\vx \in \Omega$ corresponds to a particular location in physical space.
The two are related by a bijective mapping $\bm \chi[t] : \Omega_0  \to \Omega $, such that the location in physical space of material point $\bm \xi$ at time $t$ is $\vx(\bm \xi,t) = \bm \chi[t](\bm \xi)$.
It is convenient in the case of unstressed initial conditions to let the reference frame be the initial body configuration, $\bm \chi[0](\bm \xi)  =\bm \xi$.
We will use this assumption going forward.

The deformation gradient tensor $\ten F$ is defined so that in indical notation (and dropping for the moment the notion of time-dependence) $F_{i\alpha} = \p\chi_i/\p\xi_\alpha$, where Latin and Greek indices denote vector components in the deformed and reference frames, respectively.
Physically, $\ten F$ represents the mapping of material line segments from the reference frame to the deformed frame, such that if $\bm{d\chi} \in \Omega_0$ and $\bm{dx} \in \Omega$ refer to the same material line segment in the two frames, they are related by $dx_i = F_{i\alpha} d\chi_\alpha$.
This quantity contains all of the information relating to deformation of an elastic body.
In particular, the squared principal stretches $\lambda_i^2$ are given by the eigenvalues of the left and right Cauchy-Green tensors $\ten B = \ten F \cdot \ten F^\trans$, $\ten C = \ten F^\trans \cdot \ten F$.
In general we can write $\ts = \ts(\ten F)$ for a number of widely used finite-strain models for elasticity.

To employ the RMT, we consider the reference map variable $\vX = \vX(\vx,t) = \bm \chi[t]^{-1}(\vx)$ given by the inverse of the bijective mapping $\bm \chi$.
By tracking this field relative to the constraint $D^s \vX/Dt = 0$, at each step and position in space it is possible to calculate $\ten F^{-1}$ using our finite difference stencil above, since $F^{-1}_{\alpha i} = \p X_\alpha/\p x_i$.
At the cost of an additional matrix inversion for each calculation of the elastic stress tensor, we gain the ability to simulate a wide variety of finite strain models for elastic stress on the same Eulerian grid as the rest of our simulated fields.
Here, we fix the simulated reference map field $\vX_{i,j}^n = \vX(\vx_{i,j},t_n)$ at cell centers and integrate the evolution equation $D^s \vX / Dt = 0$ using Heun's method,
\begin{equation}
	\frac{\vX^{n+1} - \vX^n}{\Delta t} = - 
	\frac12 \left[\left(\vvs \cdot \gr \vX\right)^n_{i,j} + \left(\vvs \cdot \gr \vX\right)^{\overline{n+1}}_{i,j}\right],
\end{equation}
where the advective derivatives are calculated using the ENO2 method.
Given the calculation of the deformation gradient as described, we let the solid stress be given by a neo-Hookean model \citep{lin2022},
\begin{equation}
	\ts_e = \kappa (J-1) \tI + \frac{G}{J^{1+2/D}}\Dev \left(\ten F \cdot \ten F^\trans\right), \qquad J = \Det\ten F,
\end{equation}
where $D$ is the dimension.
Each of the quantities in the above can be calculated from $\ten F$ as long as $\gr\vX$ is evaluated on cell walls using the finite difference rules established earlier.
Note that from $J = \Det \ten{F}$ and \eqref{eq:gel:phi} we can derive
\begin{equation}
	J = \frac{\phi_0}{\phi}, \label{eq:J_to_phi}
\end{equation}
i.e.\@ we can express the relative change of a differential volume element either through the determinant of the deformation gradient $J$ or through the inverse volume fraction $\phi$.
Specifically, if the solid contracts ($J < 0$) the amount of solid at a position in space must go up ($\phi > \phi_0$) and vice versa.

We assume there is also a dissipative stress related to the motion of the polymer, as modeled by others including \citet{radszuweit_active_2014} and \citet{weber_differential_2018}.
Introducing the deformation rate tensor $\tD = \frac12\left[\gr \vvs + (\gr \vvs)^\trans\right]$, we write
\begin{equation}
	\ts_v = \phi \eta_b \Tr \left(\tD\right) \tI + 2 \phi \eta_s \Dev\left(\tD\right). \label{eq:gel:visc}
\end{equation}
In conjunction with the elastic stress $\ts_e$, this yields a Kelvin--Voigt viscoelastic material which behaves like a fluid on short times and like a solid on long times.
The Crank--Nicolson update in \eqref{eq:gel:inter_s} is constructed using the same finite difference stencils as in the explicit updates.

It remains only to determine how we will solve the two linear systems arising at each time step. 
For consistent implementation across two and three dimensions, we solve each using PETSc, the Portable, Extensible Toolkit for Scientific Computation \citep{petsc-efficient,petsc-user-ref,petsc-web-page}.
The rest of the system is solved using custom-written C\texttt{++} code with Open MPI \citep{gabriel2004} for distributed-memory parallelization.
The code base is templated on dimension to minimize code propagation requirements and eliminate implementation delay between the two- and three-dimensional cases.

\section{Applications}
In this section, we apply to the method above to two sample systems in order to assess its performance, flexibility, and convergence.

\subsection{Method of manufactured solutions}
To verify the method's rate convergence in space and time, we apply it to a manufactured solution and report the error on several grid sizes. The manufactured solution requires three parameters $A$, $P$ and $\ttau$, assumes $\hat{\rho}_s = \hat{\rho}_f =: \rho$, and is given by
\begin{equation}
	\phi(x,y,t) = \frac12 + A \cos\varphi_x \cos \varphi_y \sin \varphi_t, \qquad  p(x,y,t) = P \cos\varphi_x \cos\varphi_y \sin\varphi_t,
\end{equation}
where for ease of notation we have introduced the phase variables
\begin{equation}
	\varphi_x = \frac{2\pi x}{L_x}, \ \varphi_y = \frac{2\pi y}{L_y}, \ \varphi_t = \frac{2\pi t}{\ttau}.
\end{equation}
This solid fraction is naturally produced by the velocity fields
\begin{equation}
	\vvs(x,y,t) = \left(\frac{-A L_x \sin\varphi_x\cos\varphi_y \cos \varphi_t}{\ttau + 2\ttau A \cos\varphi_x \cos\varphi_y \sin\varphi_t}, \qquad 
	 \frac{-A L_y \cos\varphi_x\sin\varphi_y \cos \varphi_t}{\ttau + 2\ttau A \cos\varphi_x \cos\varphi_y \sin\varphi_t}\right)^\trans,
\end{equation}
\begin{equation}
	\vvf(x,y,t) = \left(\frac{A L_x \sin\varphi_x\cos\varphi_y \cos \varphi_t}{\ttau - 2\ttau A \cos\varphi_x \cos\varphi_y \sin\varphi_t}, \qquad 
	 \frac{A L_y \cos\varphi_x\sin\varphi_y \cos \varphi_t}{\ttau - 2\ttau A \cos\varphi_x \cos\varphi_y \sin\varphi_t}\right)^\trans.
\end{equation}
Neglecting elasticity, the phase-specific body forces which will drive the system towards the manufactures solution are
\begin{subequations}
\begin{equation}
	\vf_s(x,y,t) = \rho \phi \frac{\p \vvs}{\p t} - \gr \cdot \ts_v - \Gamma(\phi) (\vvf - \vvs) + \phi\gr p, 
\end{equation}
\begin{equation}
	\vf_f(x,y,t) =(1- \phi) \rho \frac{\p \vvf}{\p t} - \Gamma(\phi)(\vvs - \vvf) + (1-\phi)\gr p.
\end{equation}
	\label{eq:mms_f}
\end{subequations}
The resulting expressions are cumbersome, but a symbolic solver such as Mathematica can be used to convert the expressions \eqref{eq:mms_f} directly to C\texttt{++} code.

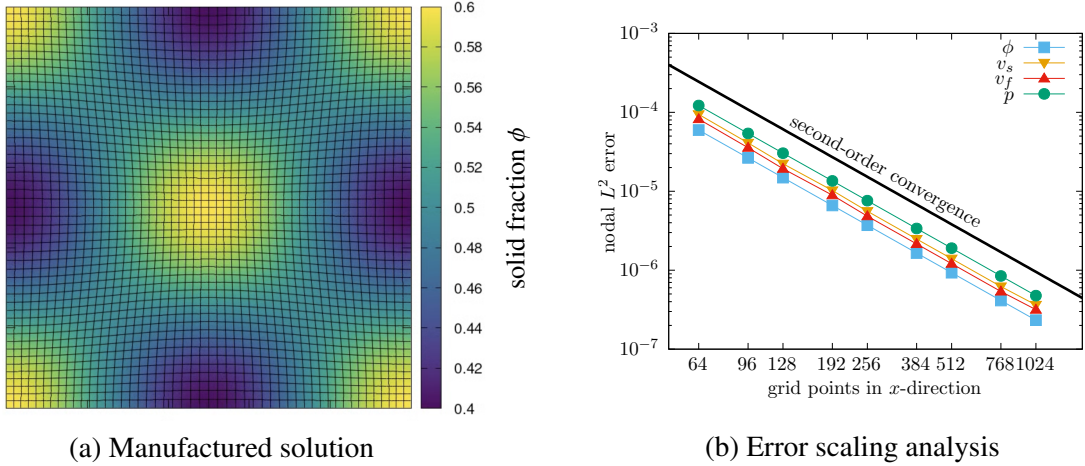
\begin{figure}
	\centering
	\begin{tikzpicture}
		\node at (-5,0) {
	\includegraphics[width=0.4\textwidth]{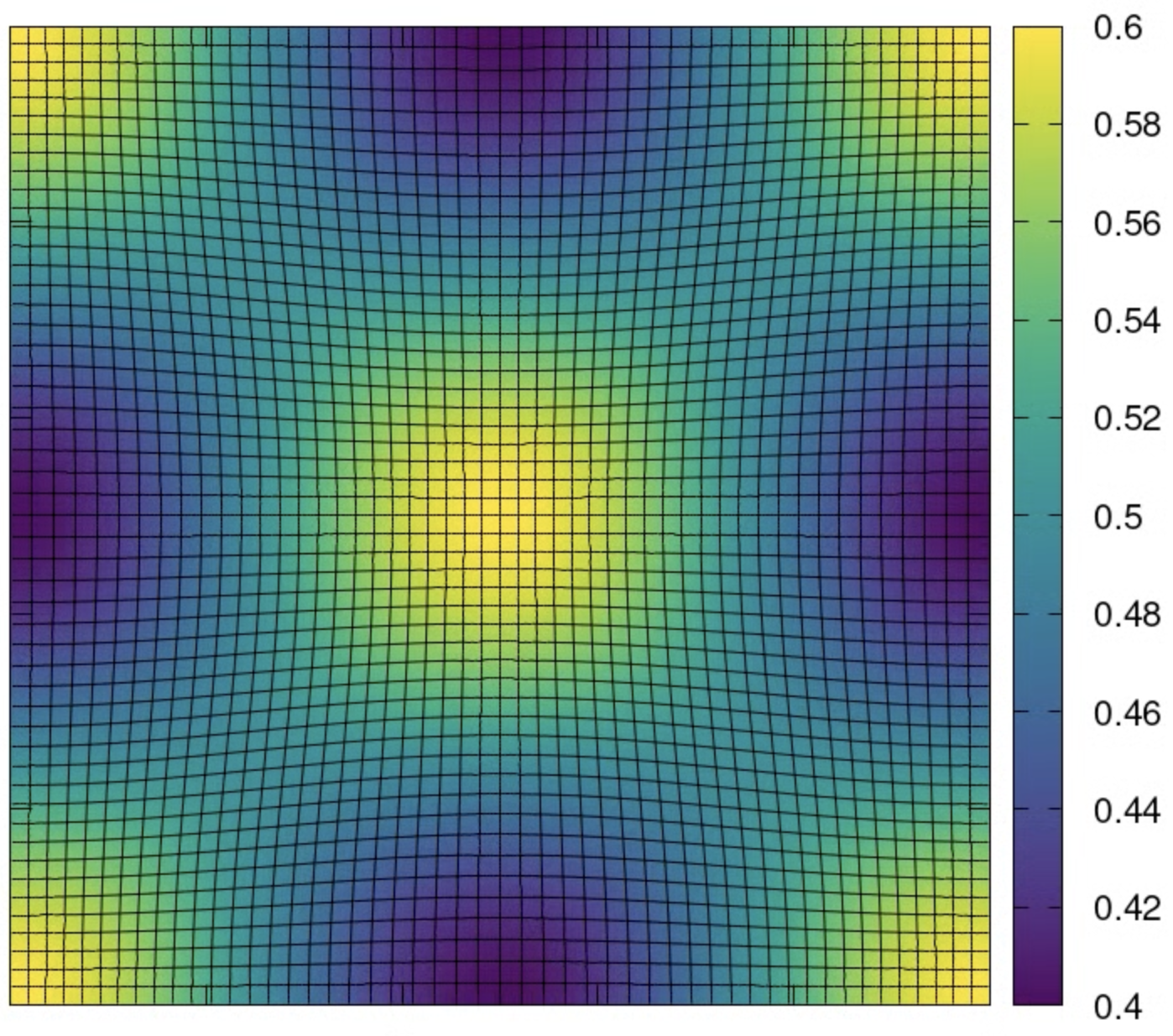}
};
		\node at (-5.3,-3.2) { \small (a) Manufactured solution};
		\node[rotate=90] at (-1.4,0) {\footnotesize solid fraction $\phi$};
		\node at (3,0) {
	\includegraphics[width=0.425\textwidth,trim=0 0 0 0,clip]{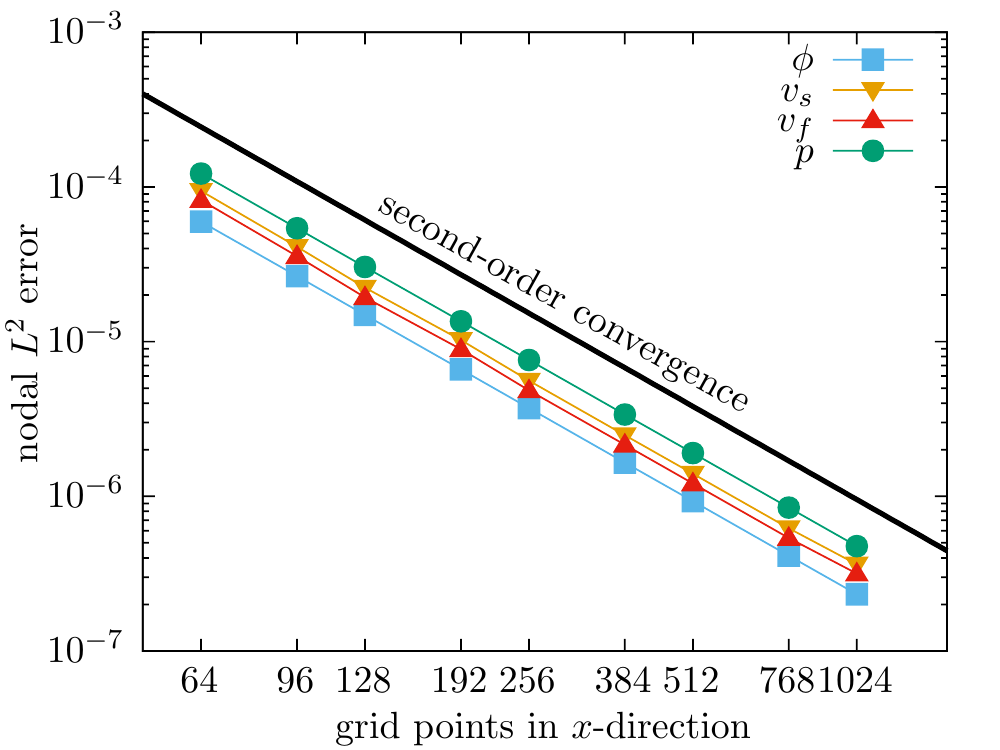}
};
	\node at (3,-3.2) { \small (b) Error scaling analysis};
	\end{tikzpicture}
    \caption{Scaling analysis using the method of manufactured solutions. (a) A known solution, shown at $t= 1.24$,  manufactured using the body forces \eqref{eq:mms_f} with $A = 0.1$, $P = 0.1$, and $\ttau = 1$ over the domain $\Omega = [0,10]^2$. The black lines indicate deformation of the solid phase from $t= 0$. (b) The system is simulated up to $t = 10$, at which point the numerical solution is compared with the actual values.  The nodal $L^2$ error is shown as a function of one-dimensional grid size, demonstrating second-order convergence in space and time. The system parameters are $\{\rho,\eta_b,\eta_s,\Gamma_0\} = \{0.01,1,0.1,1\}.$}
	\label{fig:mms}
\end{figure}

A snapshot of the solution is shown on a $256^2$ grid in \figref{fig:mms}(a). The black lines in the figure corresponds to level sets of the reference map components, and are plotted to illustrate deformation from the reference configuration. To test the convergence, the system was simulated up to $t = 10$ on square two-dimensional grids of varying sizes. The results are shown in \figref{fig:mms}(b), demonstrating second-order convergence in space and time.

\subsection{Phase separating neo-Hookean gels}
Polymer gels have been known for decades to phase separate under energetically favorable conditions \citep{flory1953}.
Typically, this is described using a free energy density of the form
\begin{equation}
	f_\text{mix} = \frac{k_B T}{\nu}\left[\phi\log\phi + (1-\phi) \log(1-\phi) + \chi \phi(1-\phi)\right],
\end{equation}
which represents the pointwise cost of maintaining a phase fraction $\phi$.
The logarithmic terms drive the system away from $\phi = 0,1$ and are derived from entropic considerations. 
The remaining term is scaled by the Flory--Huggins interaction parameter, $\chi$, a scalar that represents the polymer--solvent affinity or dis-affinity due to close-range energetic interactions.

\begin{figure}
	\centering
	\begin{tikzpicture}
		\node[anchor=south west,inner sep=0] at (0,0.01\textwidth) {
			\includegraphics[width=0.2\textwidth]{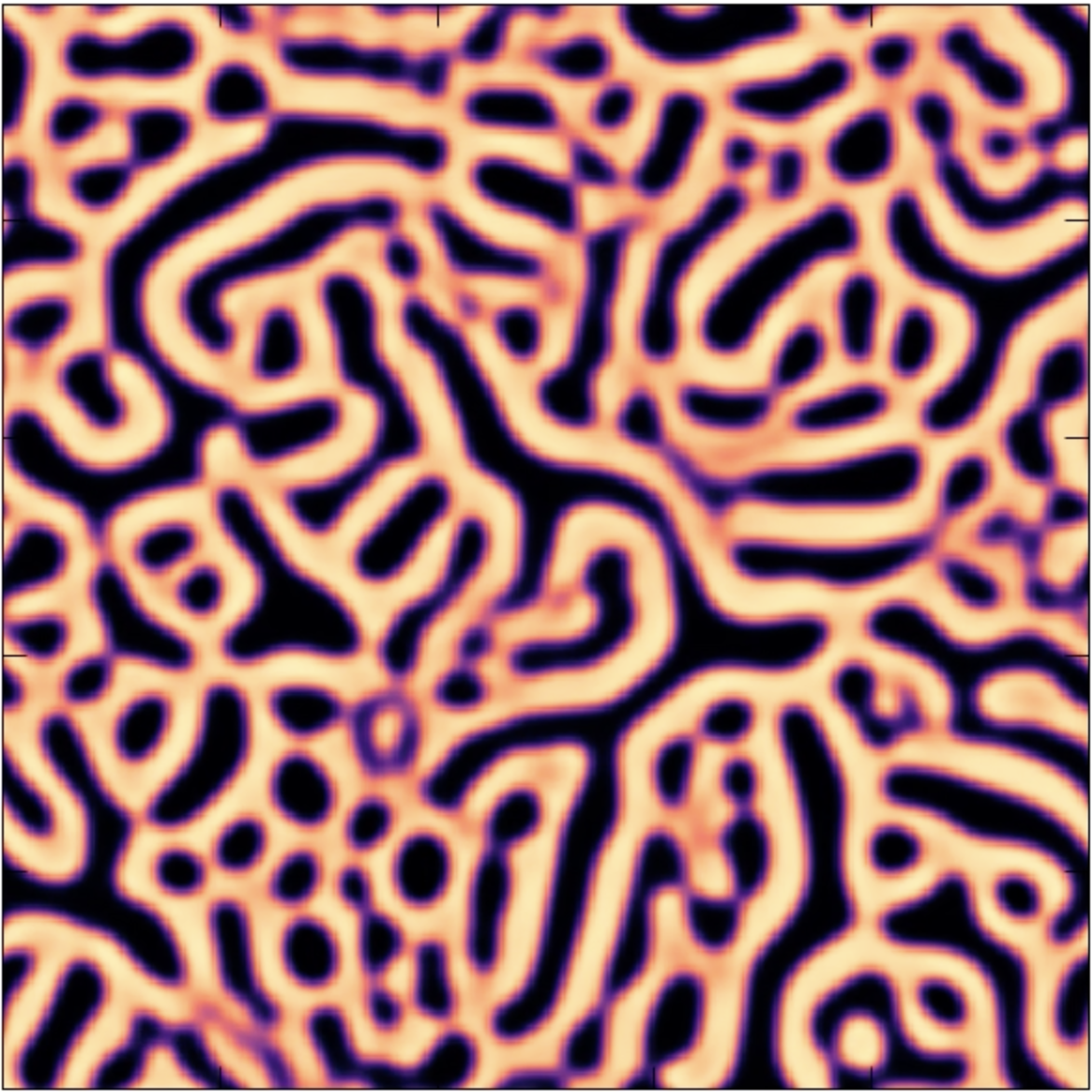}
		};
		\node[anchor=south west,inner sep=0] at (0.22\textwidth,0.01\textwidth) {
			\includegraphics[width=0.2\textwidth]{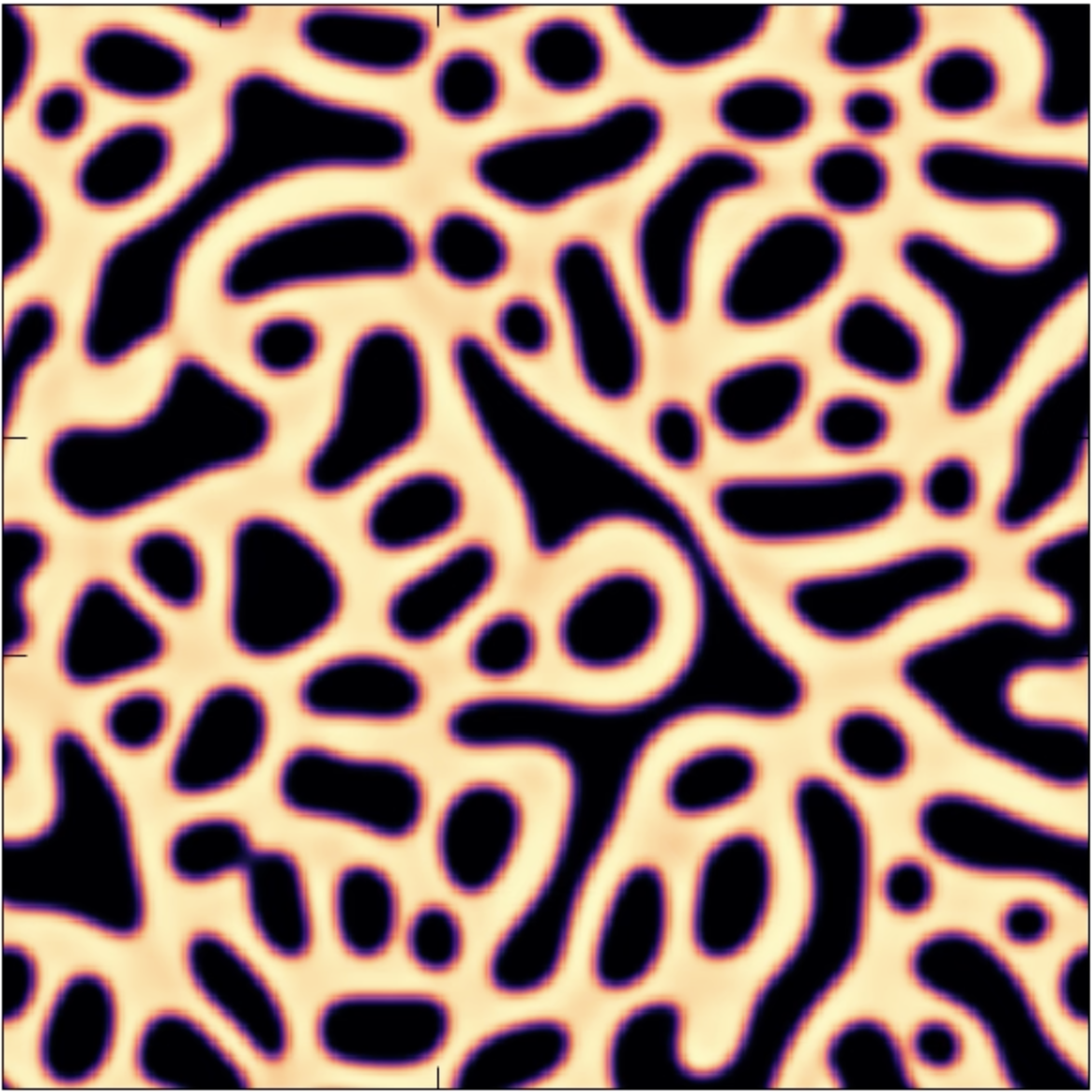}
		};
		\node[anchor=south west,inner sep=0] at (0.44\textwidth,0.01\textwidth) {
			\includegraphics[width=0.2\textwidth]{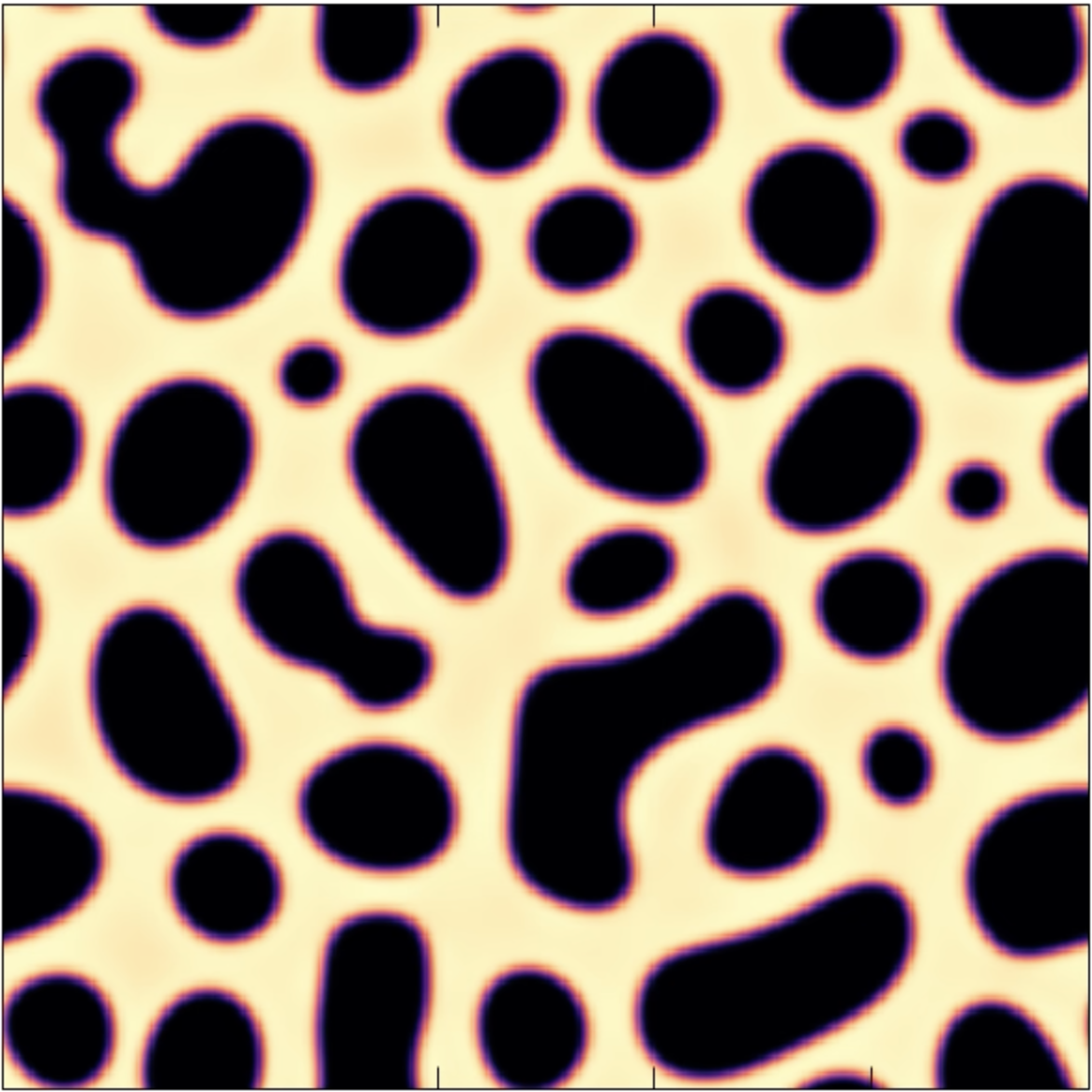}
		};
		\node[anchor=south west,inner sep=0] at (0.66\textwidth,0.01\textwidth) {
			\includegraphics[width=0.2\textwidth]{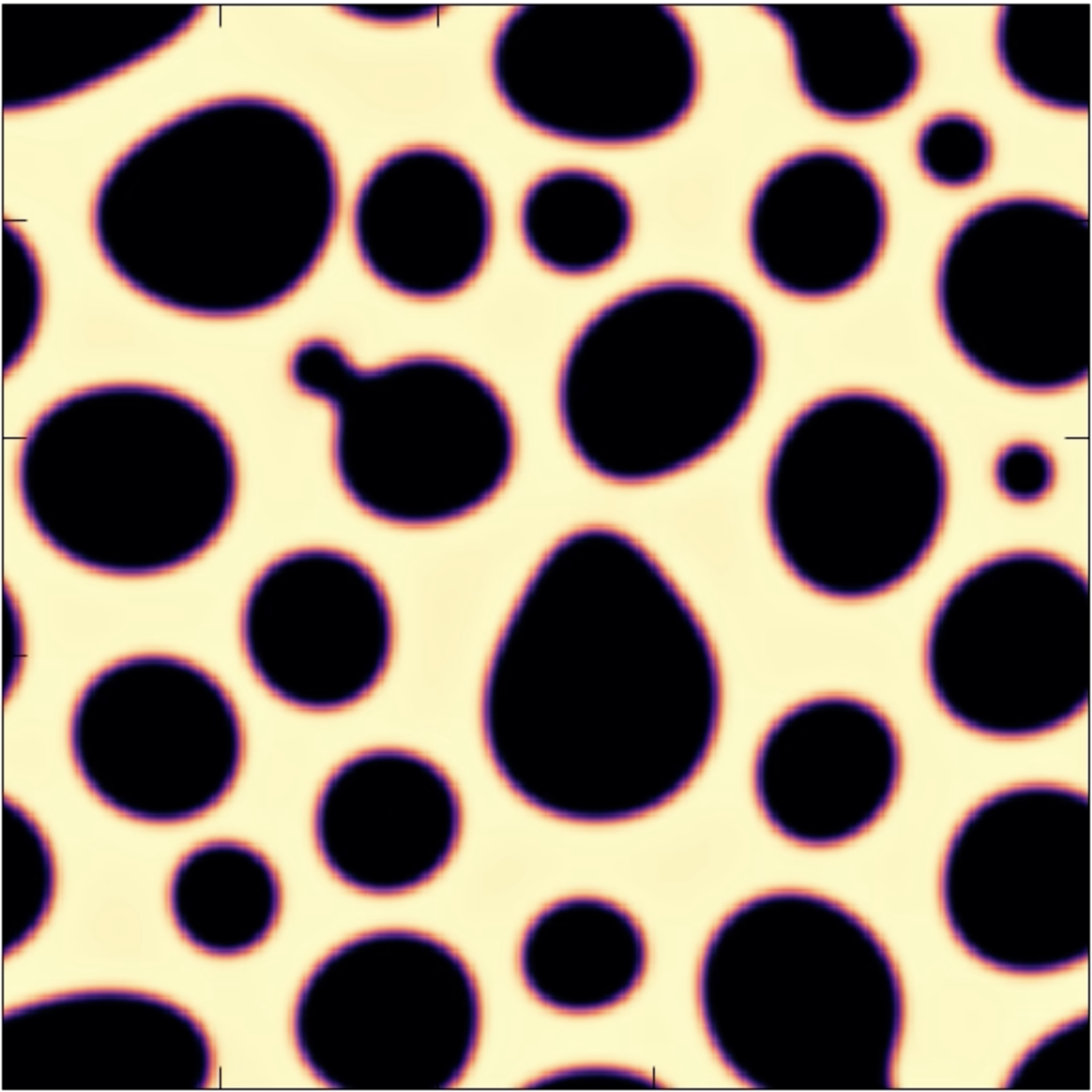}
		};

		\node[anchor=north west,inner sep=0] at (0,-0.01\textwidth) {
			\includegraphics[width=0.2\textwidth]{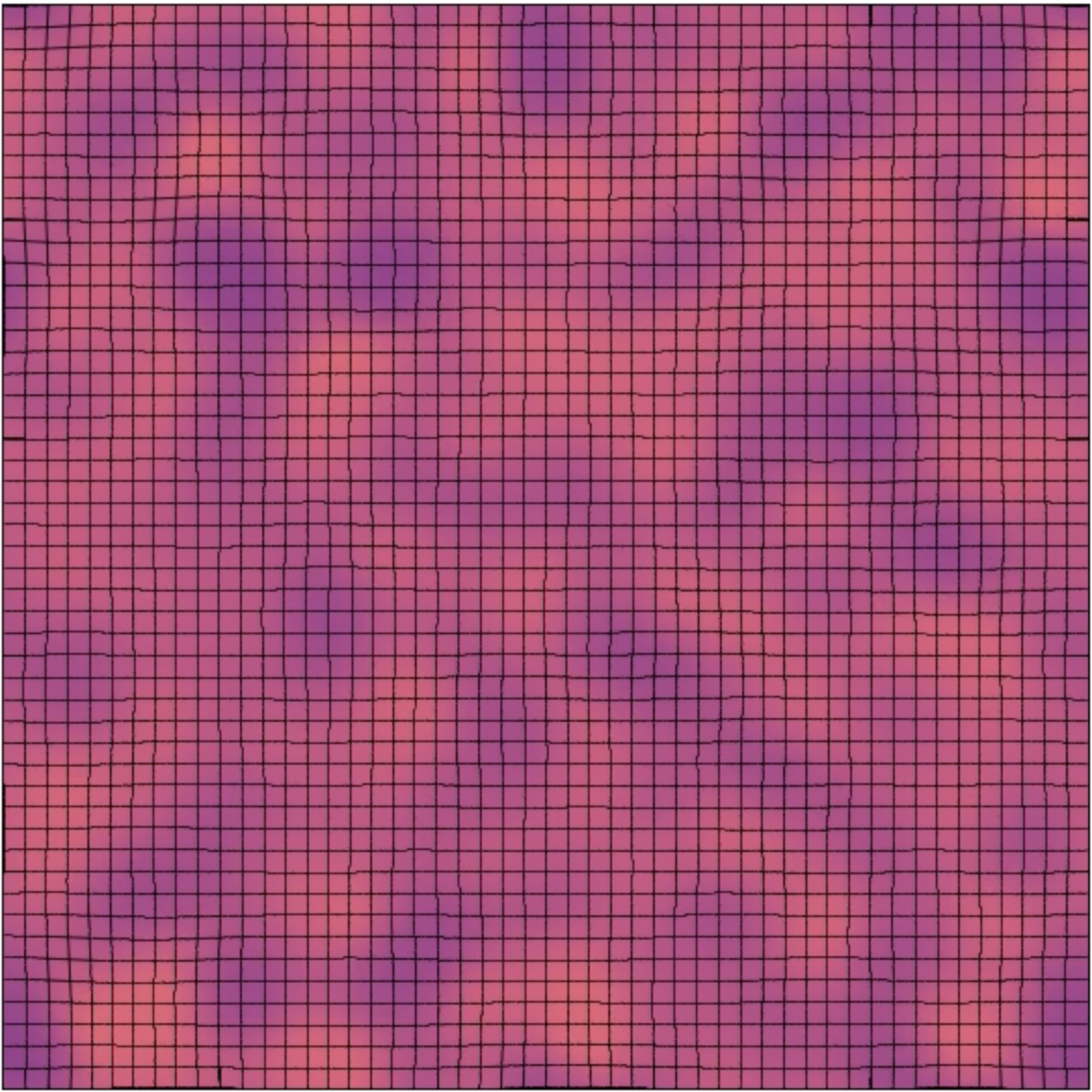}
		};
		\node[anchor=north west,inner sep=0] at (0.22\textwidth,-0.01\textwidth) {
			\includegraphics[width=0.2\textwidth]{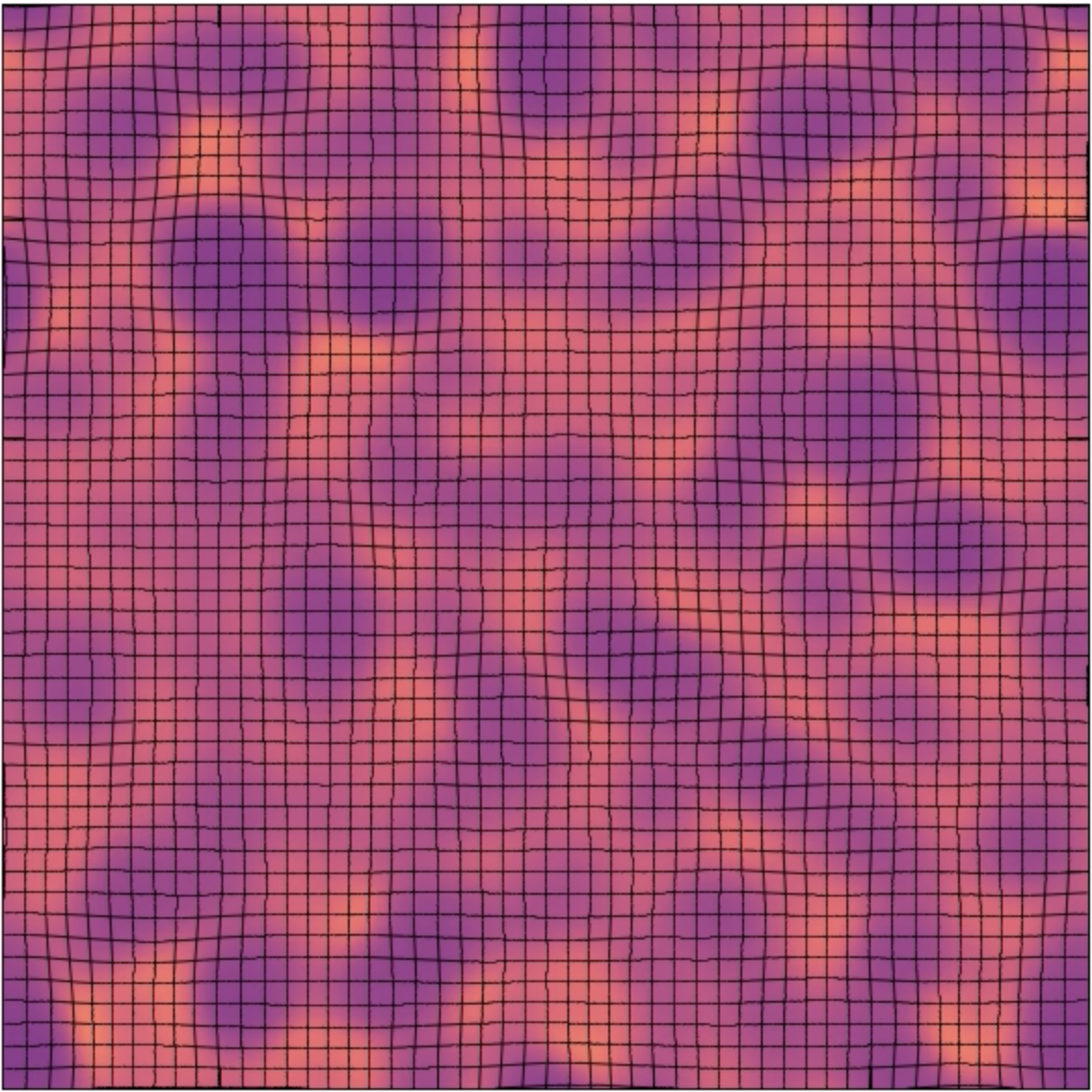}
		};
		\node[anchor=north west,inner sep=0] at (0.44\textwidth,-0.01\textwidth) {
			\includegraphics[width=0.2\textwidth]{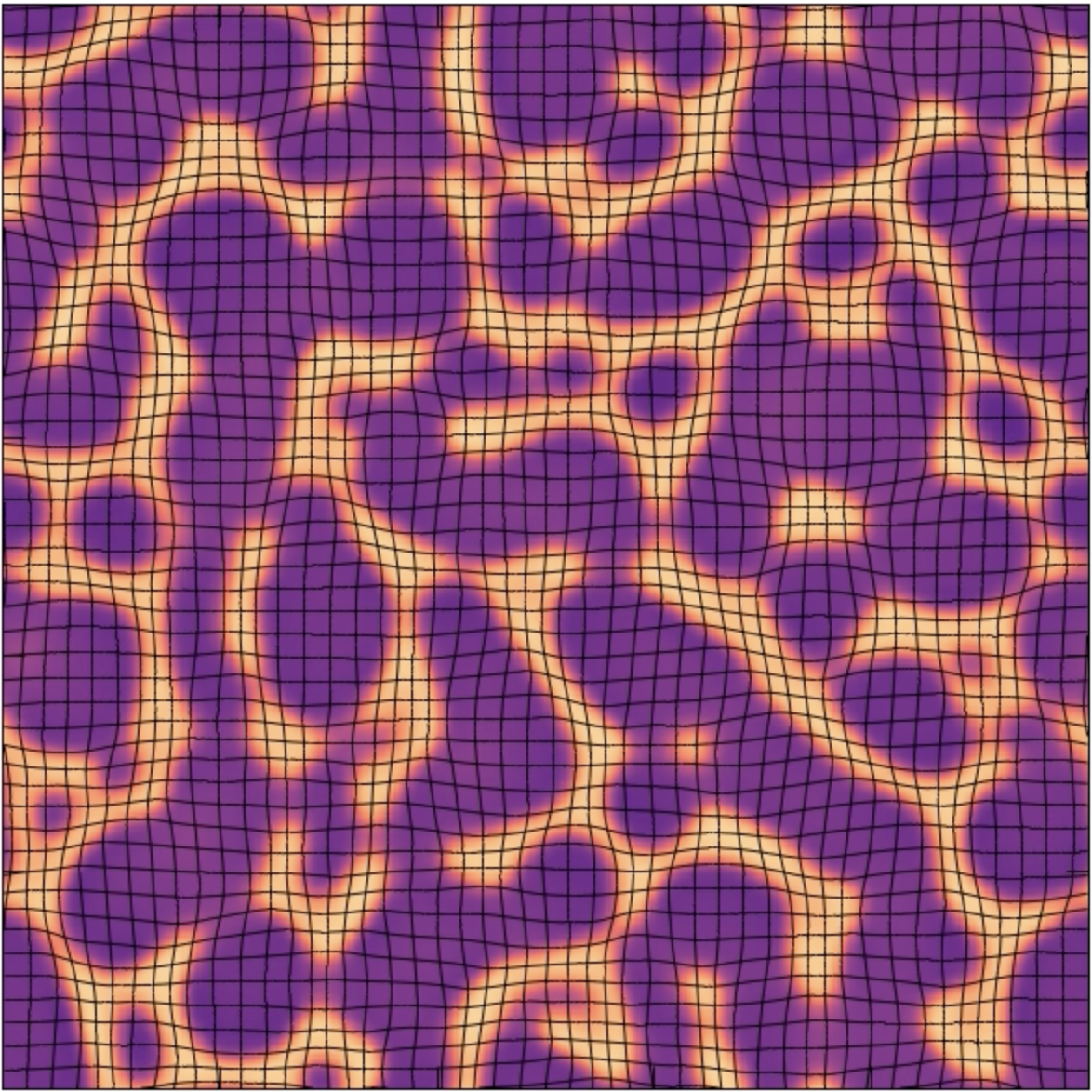}
		};
		\node[anchor=north west,inner sep=0] at (0.66\textwidth,-0.01\textwidth) {
			\includegraphics[width=0.2\textwidth]{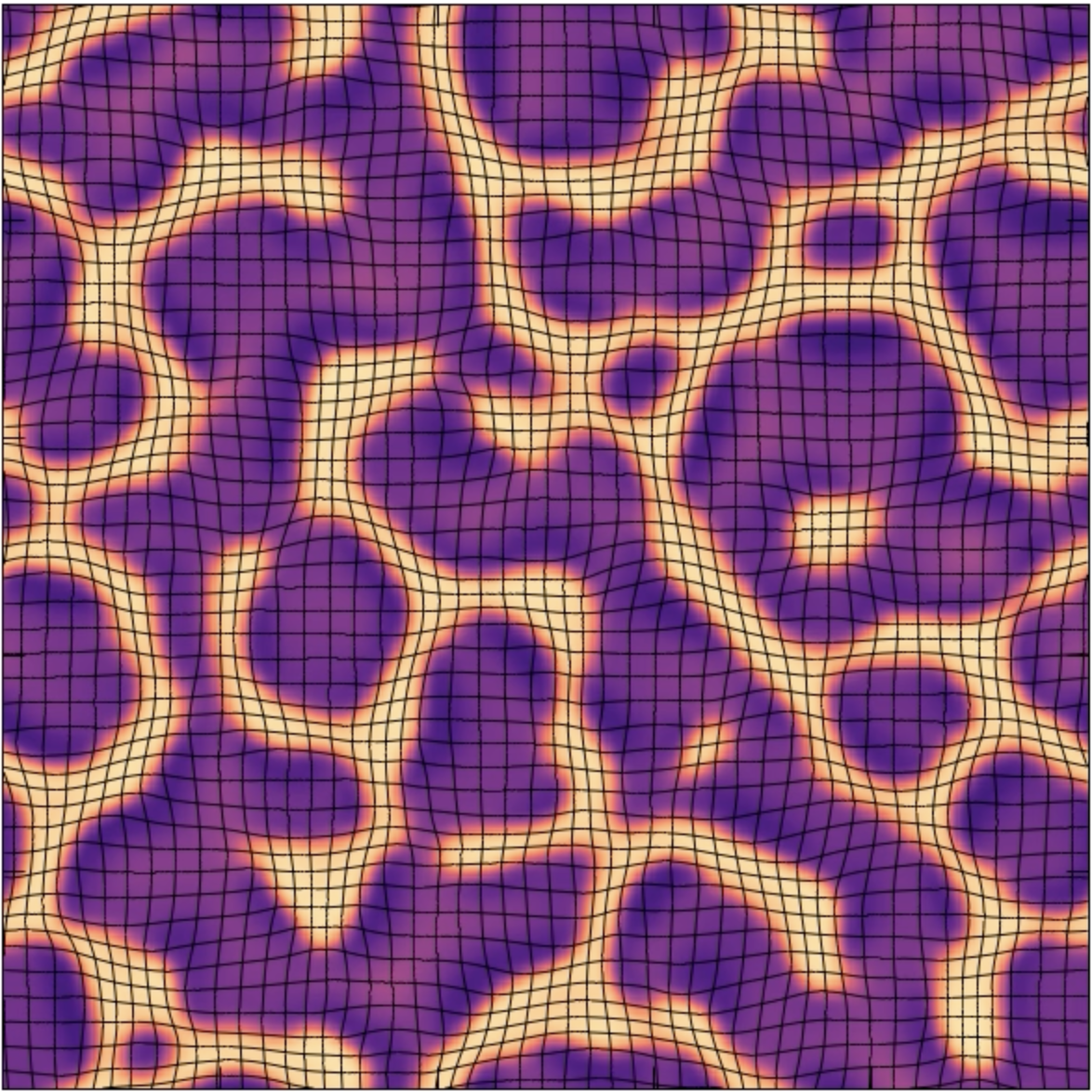}
		};

		\node[anchor=east,xshift=1em] at (0.92\textwidth,0) {
			\includegraphics[height=0.43\textwidth]{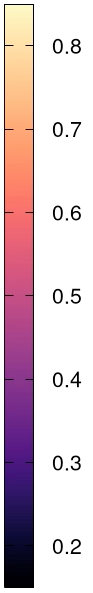}
		};
		
		\node[anchor=north,rotate=90] at (0.935\textwidth,0) {\footnotesize solid fraction $\phi$};

		\node at (0.1\textwidth,-3.65) {\footnotesize $t = 180$};
		\node at (0.32\textwidth,-3.65) {\footnotesize $t = 250$};
		\node at (0.54\textwidth,-3.65) {\footnotesize $t = 500$};
        \node at (0.76\textwidth,-3.65) {\footnotesize $t = 1000$};

		\node[rotate=90] at (-0.25,-1.7) {\footnotesize $K=0.5$, $G = 0.05$};
		\node[rotate=90] at (-0.25,1.6) {\footnotesize $K=G = 0$};

	\end{tikzpicture}
	\caption{Comparison of classical phase separation with and without elastic resistance. (a) In the absence of elasticity, a biphasic gel undergoes spinodal decomposition, forming biconnected regions and eventually isolated droplets which undergo slow ripening \citep{tanaka_unusual_1993}. (b) In the presence of elasticity, phase separation proceeds on a slower time scale and is ultimately arrested. As in \figref{fig:mms}, the black lines in the second row are level sets of the reference map field and show deformation of the solid phase. Note that while in \figref{fig:mms} the lines are purely for visualization, in this case the reference map field is used to calculate the elastic resistance opposing phase separation. The other system parameters are $\{\rho,k_bT/\nu,\eta_b,\eta_s,\Gamma_0,\chi,k\} = \{0.01,1,1,0.1,1,2.4,0.0625\}.$}
	\label{fig:comp}
\end{figure}
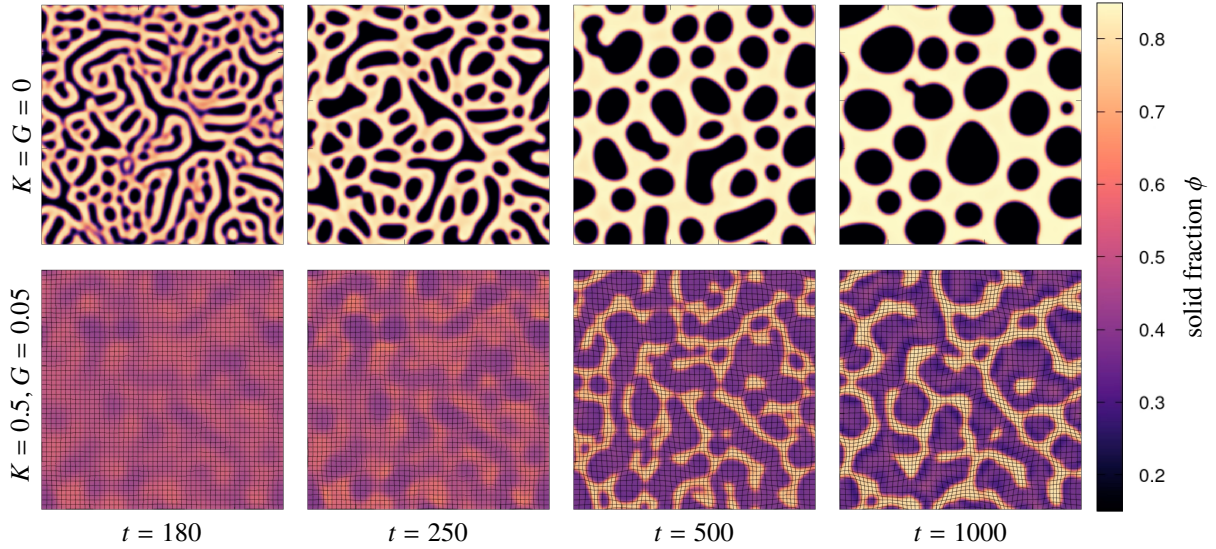

The total energy of the system is assumed to be $F = \int_V f dV$, with $f = f_\text{mix} + \frac{k}{2}|\gr \phi|^2$, such that the presence of boundaries between constant-$\phi$ phases is penalized.
This is equivalent to allowing for surface tension at the phase boundaries associated with an interface width $\sqrt{k}$.
Gradients in the chemical potential $\mu = \delta f/\delta \phi$ drive flow of the solid phase relative to the fluid in order to minimize $F$ over the system \citep{tanaka_viscoelastic_2000}.
This is represented as an additional body force $\bm f_t$ acting on the solid phase given by
\begin{equation}
	\bm f_t = -\phi \gr \frac{\p f}{\p \phi} + \phi k \gr \nabla^2 \phi,
\end{equation}
which we incorporate into our system for the remainder of this section.
We simulate two 512$^2$ systems with identical initial conditions $\phi(x,y,0) = 0.5 + \delta\phi(x,y)$, where $\delta \phi(x,y)$ is a Gaussian thermal noise field. The two systems differ only in the magnitude of the elastic stress.
Snapshots from each simulation are shown in \figref{fig:comp}.

In \figref{fig:comp}(a), the solid phase experiences no elastic stress, and we recover the behavior of liquid--liquid phase separation, albeit with asymmetric rheologies as in the work of \citet{tanaka_unusual_1993}. The initial instability proceeds unimpeded, and the gel forms biconnected regions before settling into an array of droplets which undergo slow ripening.
In \figref{fig:comp}(b), the elastic resistance slows the rate at which phase separation occurs, and eventually the dynamics are arrested as thermodynamic forces are exactly balanced by the elastic reponse to the extreme deformation.

\begin{figure}
	\centering
	\includegraphics[width=\textwidth]{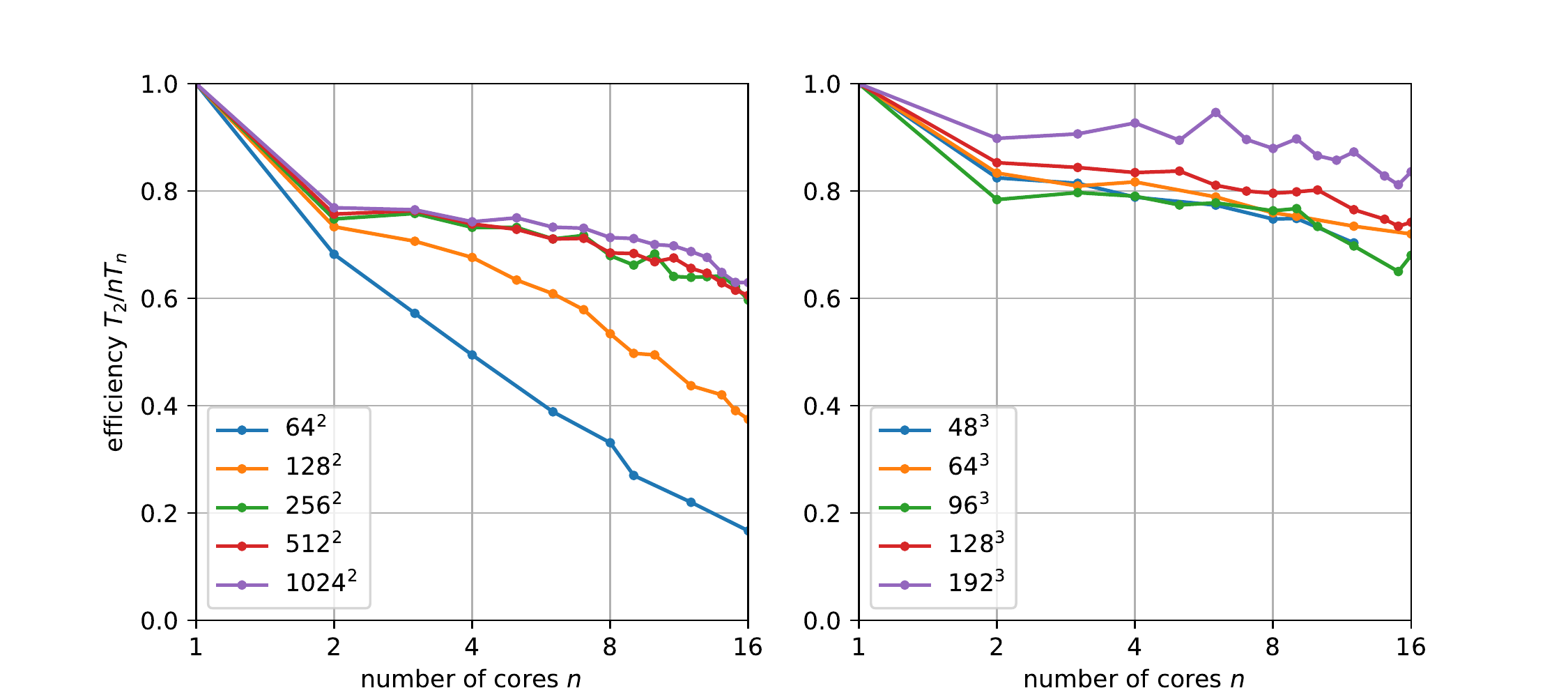}
	\caption[{Parallel efficiencies on 1--16 CPU cores}]{
		Parallel efficiencies of the computational method on 1--16 CPU cores for a variety of grid sizes in two (left) and three (right) dimensions. The duration of simulation is constant for a given grid size. The efficiency $e_k$ on $k$ cores is computed by comparing the wall time to simulate on a single core $T_1$ and on $k$ cores $T_k$. Then $e_k = T_1 / k T_k$. The domain is $[0,10]$ in each dimension, and the system parameters are $\{\rho,\kappa,G,k_bT/\nu,\eta_b,\eta_s,\Gamma_0,\chi,k\} = \{0.01,0,0,1,1,0.1,1,2.4,0.0625\}$. 
	}
	\label{fig:gel_parallel}
\end{figure}

For this system, we also report how the method scales across differing numbers of processors in two and three dimensions.
\figref{fig:gel_parallel} shows the parallel efficiency for systems of varying grid size across 1--16 CPU cores.
For a given system size, we record the serial wall clock time $T_1$ to simulate a certain number of time steps $n$.
We do not record time for initialization or data recording.
For a given number of cores $k$, we then compute the time $T_k$ to simulate $n$ timesteps and calculate the efficiency as $T_1 / kT_k$.
In two dimensions, the efficiency is not fully realized until grid sizes around 256$^2$ or above.
In that regime, it reaches around 75\% efficiency on 2 cores and decreases to 60\% on 16 cores.
The three dimensional case is generally more efficient, which is to be expected given the large system sizes.
At 2 cores, efficiencies are at or above 80\%, and at 16 cores they remain at 70--80\%.

\section{Conclusion}
In this work, we have introduced a novel numerical method for the simulation of incompressible gels which will be applied to the study of active cytoskeletal gels.
The method is completely Eulerian, leading to straightforward coupling between flow and mechanics.
Moreover, with this approach it is easy to introduce additional model fields coupled to the solid or fluid phase, allowing for the development of multi-physics methods.
It is general with respect to the solid stress and can thus be applied to linear, non-linear, elastic, and visco-elastic porestructure rheologies.

We have shown the method converges at second-order in time and space, and by using a semi-implicit update for the solid velocity phase we eliminate the second-order CFL condition which would otherwise arise due to solid viscoelastic behavior. By applying a projection step to enforce the material incompressibility condition, we avoid difficult-to-solve saddle-point structures. Similarly, we by using explicit updates and the reference map method, we allow for simulation of non-linear elastic and viscoelastic rheologies without requiring Newton solves at each time step.

Many porous media problems of interest feature boundaries, and the method is amenable to specialization towards that end through the use of level sets or the reference map itself as in \citep{lin2022}.
While the equations it simulates are often derived in the context of soil and rock mechanics, the method is well suited for describing porous systems across nature, including biological settings such as developing tissue or cellular interiors.

\section*{Acknowledgements}
We thank Christoph Weber (Max Planck Institute for the Physics of Complex Systems) and L.\@ Mahadevan (Harvard University) for useful discussions about this work. This work was supported by the National Science Foundation under Grant No.~DMS-1753203. C.~H.~R.~was partially supported by the Applied Mathematics Program of the U.S.~DOE Office of Science Advanced Scientific Computing Research under Contract No.~DE-AC02-05CH11231.
N.~J.~D.~was partially supported by the NSF--Simons Center for Mathematical and Statistical Analysis of Biology at Harvard (Award No.~1764269) and the Harvard Quantitative Biology Initiative.


\begin{thebibliography}{44}
\expandafter\ifx\csname natexlab\endcsname\relax\def\natexlab#1{#1}\fi
\providecommand{\url}[1]{\texttt{#1}}
\providecommand{\href}[2]{#2}
\providecommand{\path}[1]{#1}
\providecommand{\DOIprefix}{doi:}
\providecommand{\ArXivprefix}{arXiv:}
\providecommand{\URLprefix}{URL: }
\providecommand{\Pubmedprefix}{pmid:}
\providecommand{\doi}[1]{\href{http://dx.doi.org/#1}{\path{#1}}}
\providecommand{\Pubmed}[1]{\href{pmid:#1}{\path{#1}}}
\providecommand{\bibinfo}[2]{#2}
\ifx\xfnm\relax \def\xfnm[#1]{\unskip,\space#1}\fi
\bibitem[{Biot(1941)}]{biot1941}
\bibinfo{author}{M.~A. Biot},
\newblock \bibinfo{title}{General theory of three‐dimensional consolidation},
\newblock \bibinfo{journal}{Journal of Applied Physics} \bibinfo{volume}{12}
  (\bibinfo{year}{1941}) \bibinfo{pages}{155--164}.
\bibitem[{Rodr{\'{i}}guez-Iturbe(1997)}]{rodriguez-iturbe1997}
\bibinfo{author}{I.~Rodr{\'{i}}guez-Iturbe}, \bibinfo{title}{{Fractal River
  Basins: Change and Self-Organization}}, \bibinfo{publisher}{Cambridge
  University Press}, \bibinfo{year}{1997}.
\bibitem[{Abrams et~al.(2009)Abrams, Lobkovsky, Petroff, Straub, McElroy,
  Mohrig, Kudrolli, and Rothman}]{abrams_growth_2009}
\bibinfo{author}{D.~M. Abrams}, \bibinfo{author}{A.~E. Lobkovsky},
  \bibinfo{author}{A.~P. Petroff}, \bibinfo{author}{K.~M. Straub},
  \bibinfo{author}{B.~McElroy}, \bibinfo{author}{D.~C. Mohrig},
  \bibinfo{author}{A.~Kudrolli}, \bibinfo{author}{D.~H. Rothman},
\newblock \bibinfo{title}{Growth laws for channel networks incised by
  groundwater flow},
\newblock \bibinfo{journal}{Nature Geoscience} \bibinfo{volume}{2}
  (\bibinfo{year}{2009}) \bibinfo{pages}{193--196}.
\bibitem[{McKenzie(1984)}]{mckenzie_generation_1984}
\bibinfo{author}{D.~McKenzie},
\newblock \bibinfo{title}{The {Generation} and {Compaction} of {Partially}
  {Molten} {Rock}},
\newblock \bibinfo{journal}{Journal of Petrology} \bibinfo{volume}{25}
  (\bibinfo{year}{1984}) \bibinfo{pages}{713--765}.
\bibitem[{Spiegelman(1993)}]{spiegelman_flow_1993}
\bibinfo{author}{M.~Spiegelman},
\newblock \bibinfo{title}{Flow in deformable porous media. {Part} 1 {Simple}
  analysis},
\newblock \bibinfo{journal}{Journal of Fluid Mechanics} \bibinfo{volume}{247}
  (\bibinfo{year}{1993}) \bibinfo{pages}{17--38}.
\bibitem[{Hewitt(2011)}]{hewitt2011}
\bibinfo{author}{I.~J. Hewitt},
\newblock \bibinfo{title}{Modelling distributed and channelized subglacial
  drainage: the spacing of channels},
\newblock \bibinfo{journal}{Journal of Glaciology} \bibinfo{volume}{57}
  (\bibinfo{year}{2011}) \bibinfo{pages}{302–314}.
\bibitem[{Ehlers et~al.(2010)Ehlers, Acartürk, and
  Karajan}]{ehlers_advances_2010}
\bibinfo{author}{W.~Ehlers}, \bibinfo{author}{A.~Acartürk},
  \bibinfo{author}{N.~Karajan},
\newblock \bibinfo{title}{Advances in modelling saturated soft biological
  tissues and chemically active gels},
\newblock \bibinfo{journal}{Archive of Applied Mechanics} \bibinfo{volume}{80}
  (\bibinfo{year}{2010}) \bibinfo{pages}{467--478}.
\bibitem[{Ranft et~al.(2012)Ranft, Prost, Jülicher, and
  Joanny}]{ranft_tissue_2012}
\bibinfo{author}{J.~Ranft}, \bibinfo{author}{J.~Prost},
  \bibinfo{author}{F.~Jülicher}, \bibinfo{author}{J.~F. Joanny},
\newblock \bibinfo{title}{Tissue dynamics with permeation},
\newblock \bibinfo{journal}{The European Physical Journal E}
  \bibinfo{volume}{35} (\bibinfo{year}{2012}) \bibinfo{pages}{46}.
\bibitem[{Dembo(1989)}]{dembo_mechanics_1989}
\bibinfo{author}{M.~Dembo},
\newblock \bibinfo{title}{Mechanics and control of the cytoskeleton in {Amoeba}
  proteus},
\newblock \bibinfo{journal}{Biophysical Journal} \bibinfo{volume}{55}
  (\bibinfo{year}{1989}) \bibinfo{pages}{1053--1080}.
\bibitem[{Howard(2001)}]{howard2002}
\bibinfo{author}{J.~Howard}, \bibinfo{title}{Mechanics of Motor Proteins and
  the Cytoskeleton}, \bibinfo{publisher}{Sinauer Associates},
  \bibinfo{year}{2001}.
\bibitem[{Bendix et~al.(2008)Bendix, Koenderink, Cuvelier, Dogic, Koeleman,
  Brieher, Field, Mahadevan, and Weitz}]{bendix_quantitative_2008}
\bibinfo{author}{P.~M. Bendix}, \bibinfo{author}{G.~H. Koenderink},
  \bibinfo{author}{D.~Cuvelier}, \bibinfo{author}{Z.~Dogic},
  \bibinfo{author}{B.~N. Koeleman}, \bibinfo{author}{W.~M. Brieher},
  \bibinfo{author}{C.~M. Field}, \bibinfo{author}{L.~Mahadevan},
  \bibinfo{author}{D.~A. Weitz},
\newblock \bibinfo{title}{A {Quantitative} {Analysis} of {Contractility} in
  {Active} {Cytoskeletal} {Protein} {Networks}},
\newblock \bibinfo{journal}{Biophysical Journal} \bibinfo{volume}{94}
  (\bibinfo{year}{2008}) \bibinfo{pages}{3126--3136}.
\bibitem[{Schaller et~al.(2013)Schaller, Schmoller, Karaköse, Hammerich,
  Maier, and Bausch}]{schaller_crosslinking_2013}
\bibinfo{author}{V.~Schaller}, \bibinfo{author}{K.~M. Schmoller},
  \bibinfo{author}{E.~Karaköse}, \bibinfo{author}{B.~Hammerich},
  \bibinfo{author}{M.~Maier}, \bibinfo{author}{A.~R. Bausch},
\newblock \bibinfo{title}{Crosslinking proteins modulate the self-organization
  of driven systems},
\newblock \bibinfo{journal}{Soft Matter} \bibinfo{volume}{9}
  (\bibinfo{year}{2013}) \bibinfo{pages}{7229}.
\bibitem[{Radszuweit et~al.(2014)Radszuweit, Engel, and
  Bär}]{radszuweit_active_2014}
\bibinfo{author}{M.~Radszuweit}, \bibinfo{author}{H.~Engel},
  \bibinfo{author}{M.~Bär},
\newblock \bibinfo{title}{An {Active} {Poroelastic} {Model} for
  {Mechanochemical} {Patterns} in {Protoplasmic} {Droplets} of {Physarum}
  polycephalum},
\newblock \bibinfo{journal}{PLoS ONE} \bibinfo{volume}{9}
  (\bibinfo{year}{2014}) \bibinfo{pages}{e99220}.
\bibitem[{Rice and Cleary(1976)}]{rice1976}
\bibinfo{author}{J.~R. Rice}, \bibinfo{author}{M.~P. Cleary},
\newblock \bibinfo{title}{Some basic stress diffusion solutions for
  fluid-saturated elastic porous media with compressible constituents},
\newblock \bibinfo{journal}{Reviews of Geophysics} \bibinfo{volume}{14}
  (\bibinfo{year}{1976}) \bibinfo{pages}{227--241}.
\bibitem[{Doi(2009)}]{doi_gel_2009}
\bibinfo{author}{M.~Doi},
\newblock \bibinfo{title}{Gel {Dynamics}},
\newblock \bibinfo{journal}{Journal of the Physical Society of Japan}
  \bibinfo{volume}{78} (\bibinfo{year}{2009}) \bibinfo{pages}{052001}.
\bibitem[{Kim et~al.(2011)Kim, Tchelepi, and Juanes}]{kim2011}
\bibinfo{author}{J.~Kim}, \bibinfo{author}{H.~A. Tchelepi},
  \bibinfo{author}{R.~Juanes},
\newblock \bibinfo{title}{Stability and convergence of sequential methods for
  coupled flow and geomechanics: Fixed-stress and fixed-strain splits},
\newblock \bibinfo{journal}{Computer Methods in Applied Mechanics and
  Engineering} \bibinfo{volume}{200} (\bibinfo{year}{2011})
  \bibinfo{pages}{1591--1606}.
\bibitem[{Dana et~al.(2022)Dana, Jammoul, and Wheeler}]{dana2022}
\bibinfo{author}{S.~Dana}, \bibinfo{author}{M.~Jammoul}, \bibinfo{author}{M.~F.
  Wheeler},
\newblock \bibinfo{title}{Performance studies of the fixed stress split
  algorithm for immiscible two-phase flow coupled with linear poromechanics},
\newblock \bibinfo{journal}{Computational Geosciences} \bibinfo{volume}{26}
  (\bibinfo{year}{2022}) \bibinfo{pages}{13--27}.
\bibitem[{De~Boer(2005)}]{deboer2005}
\bibinfo{author}{R.~De~Boer}, \bibinfo{title}{Trends in Continuum Mechanics of
  Porous Media}, volume~\bibinfo{volume}{18}, \bibinfo{publisher}{Springer},
  \bibinfo{year}{2005}.
\bibitem[{Coussy et~al.(1998)Coussy, Dormieux, and
  Detournay}]{coussy_mixture_1998}
\bibinfo{author}{O.~Coussy}, \bibinfo{author}{L.~Dormieux},
  \bibinfo{author}{E.~Detournay},
\newblock \bibinfo{title}{From mixture theory to {Biot}’s approach for porous
  media},
\newblock \bibinfo{journal}{International Journal of Solids and Structures}
  \bibinfo{volume}{35} (\bibinfo{year}{1998}) \bibinfo{pages}{4619--4635}.
\bibitem[{Spiegelman and McKenzie(1987)}]{spiegelman_simple_1987}
\bibinfo{author}{M.~Spiegelman}, \bibinfo{author}{D.~McKenzie},
\newblock \bibinfo{title}{Simple 2-{D} models for melt extraction at mid-ocean
  ridges and island arcs},
\newblock \bibinfo{journal}{Earth and Planetary Science Letters}
  \bibinfo{volume}{83} (\bibinfo{year}{1987}) \bibinfo{pages}{137--152}.
\bibitem[{De~Boer and Didwania(2001)}]{deboer2001}
\bibinfo{author}{R.~De~Boer}, \bibinfo{author}{A.~K. Didwania},
\newblock \bibinfo{title}{Saturated elastic porous solids: Incompressible,
  compressible and hybrid binary models},
\newblock \bibinfo{journal}{Transport in Porous Media} \bibinfo{volume}{45}
  (\bibinfo{year}{2001}) \bibinfo{pages}{423--443}.
\bibitem[{Borrvall and Petersson(2003)}]{borrvall2003}
\bibinfo{author}{T.~Borrvall}, \bibinfo{author}{J.~Petersson},
\newblock \bibinfo{title}{Topology optimization of fluids in {Stokes} flow},
\newblock \bibinfo{journal}{International Journal for Numerical Methods in
  Fluids} \bibinfo{volume}{41} (\bibinfo{year}{2003}) \bibinfo{pages}{77--107}.
\bibitem[{Chorin(1967)}]{chorin67}
\bibinfo{author}{A.~J. Chorin},
\newblock \bibinfo{title}{A numerical method for solving incompressible viscous
  flow problems},
\newblock \bibinfo{journal}{Journal of Computational Physics}
  \bibinfo{volume}{2} (\bibinfo{year}{1967}) \bibinfo{pages}{12--26}.
\bibitem[{Chorin(1968)}]{chorin68}
\bibinfo{author}{A.~J. Chorin},
\newblock \bibinfo{title}{Numerical solution of the {N}avier--{S}tokes
  equations},
\newblock \bibinfo{journal}{Mathematics of Computation} \bibinfo{volume}{22}
  (\bibinfo{year}{1968}) \bibinfo{pages}{745--762}.
\bibitem[{Bell et~al.(1989)Bell, Colella, and Glaz}]{bell89}
\bibinfo{author}{J.~B. Bell}, \bibinfo{author}{P.~Colella},
  \bibinfo{author}{H.~M. Glaz},
\newblock \bibinfo{title}{A second-order projection method for the
  incompressible {Navier--Stokes} equations},
\newblock \bibinfo{journal}{Journal of Computational Physics}
  \bibinfo{volume}{85} (\bibinfo{year}{1989}) \bibinfo{pages}{257--283}.
\bibitem[{Almgren et~al.(1996)Almgren, Bell, and Szymczak}]{almgren96}
\bibinfo{author}{A.~S. Almgren}, \bibinfo{author}{J.~B. Bell},
  \bibinfo{author}{W.~G. Szymczak},
\newblock \bibinfo{title}{A numerical method for the incompressible
  {Navier--Stokes} equations based on an approximate projection},
\newblock \bibinfo{journal}{SIAM Journal on Scientific Computing}
  \bibinfo{volume}{17} (\bibinfo{year}{1996}) \bibinfo{pages}{358--369}.
\bibitem[{Puckett et~al.(1997)Puckett, Almgren, Bell, Marcus, and
  Rider}]{puckett97}
\bibinfo{author}{E.~G. Puckett}, \bibinfo{author}{A.~S. Almgren},
  \bibinfo{author}{J.~B. Bell}, \bibinfo{author}{D.~L. Marcus},
  \bibinfo{author}{W.~J. Rider},
\newblock \bibinfo{title}{A high-order projection method for tracking fluid
  interfaces in variable density incompressible flows},
\newblock \bibinfo{journal}{Journal of Computational Physics}
  \bibinfo{volume}{130} (\bibinfo{year}{1997}) \bibinfo{pages}{269--282}.
\bibitem[{Brown et~al.(2001)Brown, Cortez, and Minion}]{brown2001}
\bibinfo{author}{D.~L. Brown}, \bibinfo{author}{R.~Cortez},
  \bibinfo{author}{M.~L. Minion},
\newblock \bibinfo{title}{Accurate projection methods for the incompressible
  {Navier--Stokes} equations},
\newblock \bibinfo{journal}{Journal of Computational Physics}
  \bibinfo{volume}{168} (\bibinfo{year}{2001}) \bibinfo{pages}{464--499}.
\bibitem[{Weber et~al.(2018)Weber, Rycroft, and
  Mahadevan}]{weber_differential_2018}
\bibinfo{author}{C.~A. Weber}, \bibinfo{author}{C.~H. Rycroft},
  \bibinfo{author}{L.~Mahadevan},
\newblock \bibinfo{title}{Differential {Activity}-{Driven} {Instabilities} in
  {Biphasic} {Active} {Matter}},
\newblock \bibinfo{journal}{Physical Review Letters} \bibinfo{volume}{120}
  (\bibinfo{year}{2018}) \bibinfo{pages}{248003}.
\bibitem[{Carman(1937)}]{carman1937}
\bibinfo{author}{P.~C. Carman},
\newblock \bibinfo{title}{Fluid flow through granular beds},
\newblock \bibinfo{journal}{Transactions of the Institution of Chemical
  Engineers} \bibinfo{volume}{15} (\bibinfo{year}{1937})
  \bibinfo{pages}{150--166}.
\bibitem[{Kozeny(1927)}]{kozeny1927}
\bibinfo{author}{J.~Kozeny},
\newblock \bibinfo{title}{Uber kapillare leitung der wasser in boden},
\newblock \bibinfo{journal}{Sitzungsberichte der Kaiserlichen Akademie der
  Wissenschaften. Mathematisch-Naturwissenschaftliche Classe}
  \bibinfo{volume}{136} (\bibinfo{year}{1927}) \bibinfo{pages}{271--306}.
\bibitem[{Osher and Sethian(1988)}]{osher88}
\bibinfo{author}{S.~Osher}, \bibinfo{author}{J.~A. Sethian},
\newblock \bibinfo{title}{Fronts propagating with curvature-dependent speed:
  Algorithms based on {H}amilton--{J}acobi formulations},
\newblock \bibinfo{journal}{Journal of Computational Physics}
  \bibinfo{volume}{79} (\bibinfo{year}{1988}) \bibinfo{pages}{12--49}.
\bibitem[{Osher and Shu(1991)}]{osher1991}
\bibinfo{author}{S.~Osher}, \bibinfo{author}{C.-W. Shu},
\newblock \bibinfo{title}{High-order essentially nonoscillatory schemes for
  {Hamilton--Jacobi} equations},
\newblock \bibinfo{journal}{SIAM Journal on Numerical Analysis}
  \bibinfo{volume}{28} (\bibinfo{year}{1991}) \bibinfo{pages}{907--922}.
\bibitem[{Crank and Nicolson(1947)}]{crank47}
\bibinfo{author}{J.~Crank}, \bibinfo{author}{P.~Nicolson},
\newblock \bibinfo{title}{A practical method for numerical evaluation of
  solutions of partial differential equations of the heat-conduction type},
\newblock \bibinfo{journal}{Mathematical Proceedings of the Cambridge
  Philosophical Society} \bibinfo{volume}{43} (\bibinfo{year}{1947})
  \bibinfo{pages}{50--67}.
\bibitem[{Kamrin et~al.(2012)Kamrin, Rycroft, and Nave}]{kamrin12}
\bibinfo{author}{K.~Kamrin}, \bibinfo{author}{C.~H. Rycroft},
  \bibinfo{author}{J.-C. Nave},
\newblock \bibinfo{title}{Reference map technique for finite-strain elasticity
  and fluid--solid interaction},
\newblock \bibinfo{journal}{Journal of the Mechanics and Physics of Solids}
  \bibinfo{volume}{60} (\bibinfo{year}{2012}) \bibinfo{pages}{1952--1969}.
\bibitem[{Rycroft et~al.(2020)Rycroft, Wu, Yu, and Kamrin}]{rycroft20}
\bibinfo{author}{C.~H. Rycroft}, \bibinfo{author}{C.-H. Wu},
  \bibinfo{author}{Y.~Yu}, \bibinfo{author}{K.~Kamrin},
\newblock \bibinfo{title}{Reference map technique for incompressible
  fluid--structure interaction},
\newblock \bibinfo{journal}{Journal of Fluid Mechanics} \bibinfo{volume}{898}
  (\bibinfo{year}{2020}) \bibinfo{pages}{A9}.
\bibitem[{Lin et~al.(2022)Lin, Derr, and Rycroft}]{lin2022}
\bibinfo{author}{Y.~L. Lin}, \bibinfo{author}{N.~J. Derr},
  \bibinfo{author}{C.~H. Rycroft},
\newblock \bibinfo{title}{Eulerian simulation of complex suspensions and
  biolocomotion in three dimensions},
\newblock \bibinfo{journal}{Proceedings of the National Academy of Sciences}
  \bibinfo{volume}{119} (\bibinfo{year}{2022}) \bibinfo{pages}{e2105338118}.
\bibitem[{Balay et~al.(1997)Balay, Gropp, McInnes, and Smith}]{petsc-efficient}
\bibinfo{author}{S.~Balay}, \bibinfo{author}{W.~D. Gropp},
  \bibinfo{author}{L.~C. McInnes}, \bibinfo{author}{B.~F. Smith},
\newblock \bibinfo{title}{Efficient management of parallelism in object
  oriented numerical software libraries},
\newblock in: \bibinfo{editor}{E.~Arge}, \bibinfo{editor}{A.~M. Bruaset},
  \bibinfo{editor}{H.~P. Langtangen} (Eds.), \bibinfo{booktitle}{Modern
  Software Tools in Scientific Computing}, \bibinfo{publisher}{Birkh{\"{a}}user
  Press}, \bibinfo{year}{1997}, pp. \bibinfo{pages}{163--202}.
\bibitem[{Balay et~al.(2021{\natexlab{a}})Balay, Abhyankar, Adams, Benson,
  Brown, Brune, Buschelman, Constantinescu, Dalcin, Dener, Eijkhout, Gropp,
  Hapla, Isaac, Jolivet, Karpeev, Kaushik, Knepley, Kong, Kruger, May, McInnes,
  Mills, Mitchell, Munson, Roman, Rupp, Sanan, Sarich, Smith, Zampini, Zhang,
  Zhang, and Zhang}]{petsc-user-ref}
\bibinfo{author}{S.~Balay}, \bibinfo{author}{S.~Abhyankar},
  \bibinfo{author}{M.~F. Adams}, \bibinfo{author}{S.~Benson},
  \bibinfo{author}{J.~Brown}, \bibinfo{author}{P.~Brune},
  \bibinfo{author}{K.~Buschelman}, \bibinfo{author}{E.~M. Constantinescu},
  \bibinfo{author}{L.~Dalcin}, \bibinfo{author}{A.~Dener},
  \bibinfo{author}{V.~Eijkhout}, \bibinfo{author}{W.~D. Gropp},
  \bibinfo{author}{V.~Hapla}, \bibinfo{author}{T.~Isaac},
  \bibinfo{author}{P.~Jolivet}, \bibinfo{author}{D.~Karpeev},
  \bibinfo{author}{D.~Kaushik}, \bibinfo{author}{M.~G. Knepley},
  \bibinfo{author}{F.~Kong}, \bibinfo{author}{S.~Kruger},
  \bibinfo{author}{D.~A. May}, \bibinfo{author}{L.~C. McInnes},
  \bibinfo{author}{R.~T. Mills}, \bibinfo{author}{L.~Mitchell},
  \bibinfo{author}{T.~Munson}, \bibinfo{author}{J.~E. Roman},
  \bibinfo{author}{K.~Rupp}, \bibinfo{author}{P.~Sanan},
  \bibinfo{author}{J.~Sarich}, \bibinfo{author}{B.~F. Smith},
  \bibinfo{author}{S.~Zampini}, \bibinfo{author}{H.~Zhang},
  \bibinfo{author}{H.~Zhang}, \bibinfo{author}{J.~Zhang},
  \bibinfo{title}{{PETSc/TAO} Users Manual}, \bibinfo{type}{Technical Report}
  \bibinfo{number}{ANL-21/39 - Revision 3.16}, Argonne National Laboratory,
  \bibinfo{year}{2021}{\natexlab{a}}.
\bibitem[{Balay et~al.(2021{\natexlab{b}})Balay, Abhyankar, Adams, Benson,
  Brown, Brune, Buschelman, Constantinescu, Dalcin, Dener, Eijkhout, Gropp,
  Hapla, Isaac, Jolivet, Karpeev, Kaushik, Knepley, Kong, Kruger, May, McInnes,
  Mills, Mitchell, Munson, Roman, Rupp, Sanan, Sarich, Smith, Zampini, Zhang,
  Zhang, and Zhang}]{petsc-web-page}
\bibinfo{author}{S.~Balay}, \bibinfo{author}{S.~Abhyankar},
  \bibinfo{author}{M.~F. Adams}, \bibinfo{author}{S.~Benson},
  \bibinfo{author}{J.~Brown}, \bibinfo{author}{P.~Brune},
  \bibinfo{author}{K.~Buschelman}, \bibinfo{author}{E.~M. Constantinescu},
  \bibinfo{author}{L.~Dalcin}, \bibinfo{author}{A.~Dener},
  \bibinfo{author}{V.~Eijkhout}, \bibinfo{author}{W.~D. Gropp},
  \bibinfo{author}{V.~Hapla}, \bibinfo{author}{T.~Isaac},
  \bibinfo{author}{P.~Jolivet}, \bibinfo{author}{D.~Karpeev},
  \bibinfo{author}{D.~Kaushik}, \bibinfo{author}{M.~G. Knepley},
  \bibinfo{author}{F.~Kong}, \bibinfo{author}{S.~Kruger},
  \bibinfo{author}{D.~A. May}, \bibinfo{author}{L.~C. McInnes},
  \bibinfo{author}{R.~T. Mills}, \bibinfo{author}{L.~Mitchell},
  \bibinfo{author}{T.~Munson}, \bibinfo{author}{J.~E. Roman},
  \bibinfo{author}{K.~Rupp}, \bibinfo{author}{P.~Sanan},
  \bibinfo{author}{J.~Sarich}, \bibinfo{author}{B.~F. Smith},
  \bibinfo{author}{S.~Zampini}, \bibinfo{author}{H.~Zhang},
  \bibinfo{author}{H.~Zhang}, \bibinfo{author}{J.~Zhang},
  \bibinfo{title}{{PETS}c {W}eb page},
  \bibinfo{howpublished}{\url{https://petsc.org/}},
  \bibinfo{year}{2021}{\natexlab{b}}. \URLprefix \url{https://petsc.org/}.
\bibitem[{Gabriel et~al.(2004)Gabriel, Fagg, Bosilca, Angskun, Dongarra,
  Squyres, Sahay, Kambadur, Barrett, Lumsdaine et~al.}]{gabriel2004}
\bibinfo{author}{E.~Gabriel}, \bibinfo{author}{G.~E. Fagg},
  \bibinfo{author}{G.~Bosilca}, \bibinfo{author}{T.~Angskun},
  \bibinfo{author}{J.~J. Dongarra}, \bibinfo{author}{J.~M. Squyres},
  \bibinfo{author}{V.~Sahay}, \bibinfo{author}{P.~Kambadur},
  \bibinfo{author}{B.~Barrett}, \bibinfo{author}{A.~Lumsdaine}, et~al.,
\newblock \bibinfo{title}{Open {MPI}: {G}oals, concept, and design of a next
  generation {MPI} implementation},
\newblock in: \bibinfo{booktitle}{European Parallel Virtual Machine/Message
  Passing Interface Users’ Group Meeting}, \bibinfo{organization}{Springer},
  \bibinfo{year}{2004}, pp. \bibinfo{pages}{97--104}.
\bibitem[{Flory(1953)}]{flory1953}
\bibinfo{author}{P.~J. Flory}, \bibinfo{title}{Principles of Polymer
  Chemistry}, \bibinfo{publisher}{Cornell University Press},
  \bibinfo{year}{1953}.
\bibitem[{Tanaka(2000)}]{tanaka_viscoelastic_2000}
\bibinfo{author}{H.~Tanaka},
\newblock \bibinfo{title}{Viscoelastic phase separation},
\newblock \bibinfo{journal}{Journal of Physics: Condensed Matter}
  \bibinfo{volume}{12} (\bibinfo{year}{2000}) \bibinfo{pages}{R207--R264}.
\bibitem[{Tanaka(1993)}]{tanaka_unusual_1993}
\bibinfo{author}{H.~Tanaka},
\newblock \bibinfo{title}{Unusual phase separation in a polymer solution caused
  by asymmetric molecular dynamics},
\newblock \bibinfo{journal}{Physical Review Letters} \bibinfo{volume}{71}
  (\bibinfo{year}{1993}) \bibinfo{pages}{3158--3161}.

\end{thebibliography}
\end{document}